\documentclass[copyright]{eptcs}

\providecommand{\event}{MARS 2017}

\usepackage{underscore}
\usepackage{url}
\usepackage{breakurl}
\usepackage{graphicx}
\usepackage[usenames,dvipsnames]{xcolor}
\definecolor{grey}{cmyk}{0,0,0,0.8}
\hypersetup{bookmarks=true,bookmarksopen=false,colorlinks=true,linkcolor=grey,citecolor=grey,urlcolor=grey}

\newcommand{\T}[1]{\mbox{\tt #1}}
\newcommand{\CAESARADT}{C{\AE}SAR.ADT}

\title{A Large Term Rewrite System Modelling \\ a Pioneering Cryptographic Algorithm}

\def\titlerunning{A Large Term Rewrite System Modelling a Pioneering Cryptographic Algorithm}

\author{Hubert Garavel \quad\qquad Lina Marsso
   \institute{INRIA Grenoble, France}
   \institute{Univ. Grenoble Alpes, LIG, F-38000 Grenoble, France}
   \institute{CNRS, LIG, F-38000 Grenoble, France}
   \email{Hubert.Garavel@inria.fr \quad\qquad Lina.Marsso@inria.fr}
}

\def\authorrunning{H.~Garavel \& L.~Marsso}

\begin{document}
\maketitle

\begin{abstract}
	We present a term rewrite system that formally models the Message Authenticator Algorithm (MAA), which was one of the first cryptographic functions for computing a Message Authentication Code and was adopted, between 1987 and 2001, in international standards (ISO 8730 and ISO 8731-2) to ensure the authenticity and integrity of banking transactions.
	Our term rewrite system is large (13~sorts, 18 constructors, 644 non-constructors, and 684 rewrite rules), confluent, and terminating. Implementations in thirteen different languages have been automatically derived from this model and used to validate 200 official test vectors for the MAA.
\end{abstract}

\section{Introduction}

	In data security, a Message Authentication Code (MAC) is a short sequence of bits that is computed from a given message; the MAC ensures both the authenticity and integrity of the message, i.e., that the message sender is the stated one and that the message contents have not been altered.
	A MAC is more than a mere checksum, as it must be secure enough to defeat attacks; its design usually involves cryptographic keys shared between the message sender and receiver.

	One of the first MAC algorithm to gain widespread acceptance was the Message Authenticator Algorithm (also known as Message Authentication Algorithm, MAA for short) \cite{Davies-85} \cite{Davies-Clayden-88} \cite{Preneel-11} designed in 1983 by Donald Davies and David Clayden  at the National Physical Laboratory (NPL) in response to a request of the UK Bankers Automated Clearing Services.
	The MAA was adopted by ISO in 1987 and became part of the international standards 8730 \cite{ISO-8730:1990} and 8731-2 \cite{ISO-8731-2:1992}. Later, cryptanalysis of MAA revealed various weaknesses, including feasible brute-force attacks, existence of collision clusters, and key-recovery techniques \cite{Preneel-vanOorschot-96} \cite{Rijmen-Preneel-DeWin-96} \cite{Preneel-Rumen-vanOorschot-97} \cite{Preneel-vanOorschot-99} \cite{Preneel-11}. For this reason, MAA was withdrawn from ISO standards in 2002.

	From the point of view of formal methods, the MAA is interesting because of its pioneering nature, because its definition is freely available and stable, and because it is involved enough while remaining of manageable complexity. Over the past decades, various formal specifications of the MAA have been developed using VDM, Z, abstract data types (i.e., algebraic specifications), term rewrite systems, etc. For such formalisms, the usual examples often deal with syntax trees, which are explored using standard traversals (breadth-first, depth-first, etc.); contrary to such commonplace examples, cryptographic functions (and the MAA, in particular) exhibit more diverse behaviour, as they rather seek to perform irregular computations than linear ones.

	The present article is organized as follows. Section~\ref{MAA} provides an algorithmic overview of the MAA.
Section~\ref{HISTORY} lists the preexisting formal specifications of the MAA.
Section~\ref{MODEL} presents the modelling of the MAA using term rewrite systems.
Section~\ref{VALIDATION} discusses validation steps applied to this model.
Finally, Section~\ref{CONCLUSION} gives concluding remarks.

\section{Overview of the MAA}
\label{MAA}

	Nowadays, Message Authentication Codes are computed using different families of algorithms based on either cryptographic hash functions (HMAC), universal hash functions (UMAC), or block ciphers (CMAC, OMAC, PMAC, etc.). Contrary to these modern approaches, the MAA was designed as a standalone algorithm that does not rely on any preexisting hash function or cipher.

	In this section, we briefly explain the principles of the MAA. More detailed explanations can be found in \cite{Davies-85}, \cite{Davies-Clayden-88} and \cite[Algorithm 9.68]{Menezes-vanOorschot-Vanstone-96}.

	The MAA was intended to be implemented in software and to run on 32-bit computers. Hence, its design intensively relies on 32-bit words (called {\em blocks\/}) and 32-bit machine operations.

	The MAA takes as inputs a key and a message. The key has 64 bits and is split into two blocks $J$ and $K$. The message is seen as a sequence of blocks. If the number of bytes of the message is not a multiple of four, extra null bytes are added at the end of the message to complete the last block. The size of the message should be less than 1,000,000 blocks; otherwise, the MAA result is said to be undefined; we believe that this restriction, which is not inherent to the algorithm itself, was added in the ISO 8731-2 standard to provide MAA implementations with an upper bound (four megabytes) on the size of memory buffers used to store messages.

	The MAA produces as output a block, which is the MAC value computed from the key and the message. The fact that this result has only 32 bits proved to be a major weakness enabling cryptographic attacks; MAC values computed by modern algorithms now have a much larger number of bits. Apart from the aforementioned restriction on the size of messages, the MAA behaves as a totally-defined function; its result is deterministic in the sense that, given a key and a message, there is only a single MAC result, which neither depends on implementation choices nor on hidden inputs, such as nonces or randomly-generated numbers.

	The MAA calculations rely upon conventional 32-bit logical and arithmetic operations, among which: \T{AND} (conjunction), \T{OR} (disjunction), \T{XOR} (exclusive disjunction), \T{CYC} (circular rotation by one bit to the left), \T{ADD} (addition), \T{CAR} (carry bit generated by 32-bit addition), \T{MUL} (multiplication, sometimes decomposed into \T{HIGH\_MUL} and \T{LOW\_MUL}, which denote the most- and least-significant blocks in the 64-bit product of a 32-bit multiplication).
	On this basis, more involved operations are defined, among which \T{MUL1} (result of a 32-bit multiplication modulo $2^{32}-1$), \T{MUL2} (result of a 32-bit multiplication modulo $2^{32}-2$), \T{MUL2A} (faster version of \T{MUL2}), \T{FIX1} and \T{FIX2} (two unary functions\footnote{The names \T{FIX1} and \T{FIX2} are borrowed from \cite[pages 36 and 77]{Munster-91-a}.} respectively defined as $x \rightarrow \T{AND} (\T{OR} (x, \T{A}), \T{C})$ and $x \rightarrow \T{AND} (\T{OR} (x, \T{B}), \T{D})$, where \T{A}, \T{B}, \T{C}, and \T{D} are four hexadecimal block constants \T{A} = 02040801, \T{B} = 00804021, \T{C} = BFEF7FDF, and \T{D} = 7DFEFBFF).
	The MAA operates in three successive phases:

\begin{itemize}
	\item The {\em prelude\/} takes the two blocks $J$ and $K$ of the key and converts them into six blocks $X_0$, $Y_0$, $V_0$, $W$, $S$, and $T$. This phase is executed once. After the prelude, $J$ and $K$ are no longer used.

	\item The {\em main loop\/} successively iterates on each block of the message. This phase maintains three variables $X$, $Y$, and $V$ (initialized to $X_0$, $Y_0$, and $V_0$, respectively), which are modified at each iteration. The main loop also uses the value of $W$, but neither $S$ nor $T$.

	\item The {\em coda\/} adds the blocks $S$ and $T$ at the end of the message and performs two more iterations on these blocks. After the last iteration, the MAA result is $\T{XOR} (X, Y)$.
\end{itemize}

	In 1987, the ISO 8731-2 standard \cite[Section~5]{ISO-8731-2:1987} introduced an additional feature (called {\em mode of operation\/}), which concerns messages longer than 256 blocks and which, seemingly, was not present in the early MAA versions designed at NPL. Each message longer than 256 blocks must be split into {\em segments\/} of 256 blocks each, with the last segment possibly containing less than 256 blocks.
	The above MAA algorithm (prelude, main loop, and coda) is applied to the first segment, resulting in a value noted $Z_1$. This block $Z_1$ is then inserted before the first block of the second segment, leading to a 257-block message to which the MAA algorithm is applied, resulting in a value noted $Z_2$. This is done repeatedly for all the $n$ segments, the MAA result $Z_i$ computed for the $i$-th segment being inserted before the first block of the $(i+1)$-th segment. Finally, the MAC for the entire message is the MAA result $Z_n$ computed for the last segment.

\section{Prior Formal Specifications of the MAA}
\label{HISTORY}

	The informal description of the MAA can be found both in ISO standard 8731-2 \cite{ISO-8731-2:1992} or in a 1988 NPL technical report \cite{Davies-Clayden-88}. On this basis, several formal models of the MAA have been developed:

\begin{itemize}
	\item In 1990, G. I. Parkin and G. O'Neill designed a formal specification of the MAA in VDM \cite{Parkin-ONeill-90} \cite{Parkin-ONeill-91}. To our knowledge, this was the first attempt at applying formal methods to the MAA. The VDM specification became part of the ISO standard defining the MAA \cite[Annex~B]{ISO-8731-2:1992}.
	Three implementations in C \cite[Annex~C]{Parkin-ONeill-90}, Miranda \cite[Annex~B]{Parkin-ONeill-90}, and Modula-2 \cite{Lampard-91} were written by hand along the lines of the VDM specification.

	\item In 1991, M. K. F. Lai formally described the MAA using the set-theoretic Z notation \cite{Lai-91}. He adopted Knuth's ``literate programming'' approach, by inserting formal fragments of Z~code in the natural-language description of the MAA.

	\item In 1991, Harold~B. Munster produced a formal specification of the MAA in LOTOS \cite{Munster-91-a}. The MAA was described using only the data part of LOTOS: the behavioural part of LOTOS, which serves to describe concurrent processes, was not used. The LOTOS specification, which made intensive use of the predefined LOTOS data-type libraries, was mainly declarative but not executable, as all facilities of LOTOS abstract data types were used in an  unconstrained way. For instance, some equations could be rephrased as: ``{\em given $x$, the result is $y$ such that $x = f (y)$}'', which required to invert function $f$ in order to compute $y$.

	\item In 1992, Hubert Garavel and Philippe Turlier, taking the aforementioned LOTOS specification as a starting point, gradually transformed it by successive modifications. Their goal was to obtain an executable specification that could be processed by the \CAESARADT\ compiler \cite{Garavel-89-c} \cite{Garavel-Turlier-93}, while staying as close as possible to the original LOTOS specification.
	To do so, three main kinds of modifications were applied: (i) the LOTOS algebraic equations, which are not oriented, were turned into rewrite rules, which are oriented from left to right and, thus, more amenable to automatic execution; (ii) a distinction was made between constructor and non-constructor operations, and the discipline of ``free'' constructors was enforced --- namely, each rule defining a non-constructor $f$ must have the form ``$f (P_1, ..., P_n) \rightarrow ...$'', where each $P_i$ contains only constructors and free variables; (iii) some LOTOS sorts and operations were implemented as C types and functions, by importing manually-written C~code --- for instance, addition, multiplication, and bit shifts on 32-bit words were implemented directly in C.
        From this specification, the \CAESARADT\ compiler could automatically generate C~code that, combined with a small handwritten main program, computed the MAC value corresponding to a message and a key.
\end{itemize}

\section{Specification of the MAA as a Term Rewrite System}
\label{MODEL}

	Taking Garavel~\&~Turlier's LOTOS specification as a starting point, our central contribution is a formal model of the MAA specified as a term rewrite system.
	This model is expressed using the notations of the simple rewriting language REC proposed in \cite[Sect.~3]{Duran-Roldan-Balland-et-al-09} and \cite[Sect.~3.1]{Duran-Roldan-Bach-10}, which was lightly enhanced to distinguish between free constructors (declared in the ``\T{CONS}'' part) and non-constructors (declared in the ``\T{OPNS}'' part and defined by rewrite rules given in the ``\T{RULES}'' part).

	The model is given in Annex~\ref{ANNEX-REC} of the present article.
	Notice that the model has a few (only six) conditional rules and is thus a conditional term rewrite system; if needed, it could easily be turned into a non-conditional term rewrite system by slightly modifying the definitions of three functions (i.e., adding extra parameters and auxiliary functions), as explained in Annex~\ref{ANNEX-REC}. Our main results are the following:

\begin{itemize}
	\item Our model is {\em large}. It is 1575-line long and contains 13~sorts, 18~constructors, 644 non-constructors, and 684 rewrite rules. Although research on term rewriting led to a wealth of scientific publications, it is difficult to find concrete examples of large term-rewrite systems: for instance, in the data base of models accumulated during the three Rewrite Engines Competition (2006, 2008, and 2010), the largest models are less than 300-line long. There exist indeed larger (e.g., 10,000-line long) term rewrite systems, but they are either generated automatically (and, thus, difficult to understand by humans) or they are actual implementations of compilers or translators (and, often, are not ``pure'' term rewrite systems, as they rely upon higher-level features, e.g., subsorts or strategies).

	\item Our model is {\em exhaustive}, as it fully describes the MAA algorithm, including its ``mode of operation'' and its segmentation of messages larger than 1024 bytes.

	\item Our model is {\em self-contained}, as each detail of the MAA is expressed using term rewrite systems only; the model does not rely upon any externally-defined type or function and is thus independent from machine-specific assumptions, e.g., 32-bit vs 64-bit words or little- vs big-endian ordering.

	\item Our model is {\em executable}. From a theoretical point of view, this was enabled by the aforementioned shift from general LOTOS abstract data types to term rewrite systems, which are less declarative and more operational. From a practical point of view, this shift was not sufficient, as the MAA intensively manipulates block values (i.e., 32-bit numbers), which cannot be reasonably implemented in the Peano style (the execution stack quickly overflows when these numbers are represented using the \T{zero} and \T{succ} constructors).
	To overcome this issue, we chose to represent blocks in binary form, as words of four octets.
	So doing, the logical operations on blocks (\T{AND}, \T{OR}, \T{XOR}, and \T{CYC}) are easy to define using bitwise and octetwise manipulations. The arithmetical operations (\T{ADD}, \T{CAR}, and \T{MUL}) are more involved: we implemented them using 8-bit, 16-bit, and 32-bit adders and multipliers, more or less inspired from the theory of digital circuits.

	\item Our model is {\em minimal}, in the sense that each sort, constructor, and non-constructor defined in our model is actually used (i.e., the model does not contain ``dead'' code).

	\item Our model is {\em readable}. Despite its size, efforts have been made to give it a modular structure, which is reflected in the sections of Annex~\ref{ANNEX-REC}. Particular care has been taken to choose constructors appropriately and to keep non-constructor definitions as simple as possible.

\end{itemize}

\newpage
\section{Validation of the MAA Model}
\label{VALIDATION}

	In this section, we detail the various steps performed to make sure that our model is correct:

\begin{itemize}
	\item Our model is {\em self-checking}. Because the REC language has no input/output primitive and no provision for interfacing with external C~code, it cannot be used to compute the MAC value of a given file. In order to check whether our model was correct or not, we enriched it with 203~test vectors originating from three sources, namely:
	(i) all the test vectors provided in Tables~1 to~6 of \cite[Annex~A]{ISO-8731-2:1992} and \cite{Davies-Clayden-88};
	(ii) all the test vectors provided in \cite[Annex~E.3.3]{ISO-8730:1990} --- the subsequent test vectors of \cite[Annexes~E.3.4 and~E.4]{ISO-8730:1990} were discarded because of their size (they deal with two messages having 84 and 588 blocks, which would have led to a much too large REC specification);
	(iii) supplementary test vectors intended to specifically check for certain aspects (byte permutations and message segmentation) that were not enough covered by the above tests; this was done by introducing a \T{makeMessage} function acting as a pseudo-random message generator (see Annex~\ref{ANNEX-REC-MESSAGE}).

	\item Our model is {\em confluent}. This is easy to see, because all constructors are free and all the rules defining non-constructors have disjoint patterns and mutually exclusive premises; for safety, the disjunction of patterns and exclusion of guards has been checked mechanically by translating the REC specification to the Opal language \cite{Didrich-Fett-Gerke-et-al-94}, whose compiler emits warnings in presence of ``ambiguous'' (i.e., nondeterministic) rules.

	\item Our model is {\em terminating}. This has been verified by automatically translating our REC model into the input formalism TRS of the AProVE tool \cite{Giesl-Brockschmidt-Emmes-et-al-14}, which produced a proof of quasi-decreasingness in 76~steps.

	\item Our model was {\em validated}. We checked that the REC specification satisfies all the aforementioned test vectors.
	Because it enjoys the confluence and termination properties, all rewrite strategies should lead to the same result. Using a software framework\footnote{\url{http://gforge.inria.fr/scm/viewvc.php/rec/2015-CONVECS}} under development at INRIA Grenoble, we automatically translated our REC model into thirteen different languages: Clean, Haskell, LNT, LOTOS, Maude, mCRL2, OCaml, Opal, Rascal, Scala, SML, Stratego/XT, and Tom. We submitted these translations to sixteen compilers, interpreters, and rewrite engines:
	eleven of them reported that all the~203 tests passed successfully, while the other tools halted or timed out.
	Moreover, some involved components (namely, the binary adders and multipliers) have been validated separately using more than 30,000 test vectors.

\end{itemize}

\section{Conclusion}
\label{CONCLUSION}

	Twenty-five years after, we revisited the Message Authenticator Algorithm (MAA), which used to be a pioneering case study for cryptography in the 80s and for formal methods in the early 90s.

	We developed a formal specification of the MAA, expressed as a term rewrite system encoded in the REC language.
	As far as we are aware, it is one of the largest handwritten term rewrite systems publicly available.
	This specification is self-contained and self-checking, as it includes 203~test vectors.
	It has been carefully validated using a dozen tools.

	Parts of this specification (in particular, the binary adders and multipliers) are certainly reusable for different purposes, e.g., formal libraries for modular arithmetic or cryptography.

	This study enabled us to discover various mistakes in prior MAA specifications. For instance, we corrected the test vectors given for function PAT at the bottom of Table~3 in \cite[Annex~A]{ISO-8731-2:1992} and \cite{Davies-Clayden-88} (see Annex~\ref{ERRATA-8731} of the present article for details).
	We also corrected the handwritten implementation in~C of the function \T{HIGH_MUL} imported by the aforementioned LOTOS specification (this illustrates the risks arising when formal and non-formal codes are mixed).

	It is however fair to warn the reader that term rewrite systems are a low-level theoretical model that does not scale well to large problems. The REC specification is between two and six times longer than any other (formal or informal) description of the MAA, and it took considerable effort to come up with a REC specification that is readable, properly structured, and seemingly straightforward. Similar results might not be easy to reproduce on a regular basis with other case studies.

\subsection*{Acknowledgements}

	We are grateful to Keith Lockstone for his advices and his web site\footnote{\url{http://www.cix.co.uk/~klockstone}} giving useful information about the MAA, and to Sharon Wilson, librarian of the National Physical Laboratory, who provided us with valuable early NPL reports that cannot be fetched from the web.

\appendix

\section{Errata Concerning Annex A of the ISO-8731-2:1992 Standard}
\label{ERRATA-8731}

	After checking carefully all the test vectors contained in the original NPL report defining the MAA \cite{Davies-Clayden-88} and in the 1992 version of the MAA standard \cite{ISO-8731-2:1992}, we believe that there are mistakes\footnote{We used the French version of this standard, but we believe that the language plays no role, as the same mistakes were already present in the 1988 NPL report.} in the test vectors given for function \T{PAT}.

	More precisely, the three last lines of Table~3 \cite[page 15]{Davies-Clayden-88} --- identically reproduced in Table~A.3 of \cite[Sect.~A.4]{ISO-8731-2:1992} --- are written as follows:

\begin{small}
\begin{verbatim}
	{X0,Y0}     0103 0703 1D3B 7760     PAT{X0,Y0} EE
	{V0,W}      0103 050B 1706 5DBB     PAT{V0,W}  BB
	{S,T}       0103 0705 8039 7302     PAT{S,T}   E6
\end{verbatim}
\end{small}

	Actually, the inputs of function \T{PAT} should not be \verb+{X0,Y0}+, \verb+{V0,W}+, \verb+{S,T}+ but rather \verb+{H4,H5}+, \verb+{H6,H7}+, \verb+{H8,H9}+, the values of \T{H4}, ..., \T{H9} being those listed above in Table~3. Notice that the confusion was probably caused by the following algebraic identities:

\begin{small}
\begin{verbatim}
	{X0,Y0} = BYT (H4, H5)
	{V0,W}  = BYT (H6, H7)
	{S,T}   = BYT (H8, H9)
\end{verbatim}
\end{small}

	If one gives \verb+{X0,Y0}+, \verb+{V0,W}+, \verb+{S,T}+ as inputs to \T{PAT}, then the three results of \T{PAT} are equal to \T{00} and thus cannot be equal to \T{EE}, \T{BB}, \T{E6}, respectively.

	But if one gives \verb+{H4,H5}+, \verb+{H6,H7}+, \verb+{H8,H9}+ as inputs to \T{PAT}, then the results of \T{PAT} are the expected values \T{EE}, \T{BB}, \T{E6}.

	Thus, we believe that the three last lines of Table 3 should be modified as follows:

\begin{small}
\begin{verbatim}
	{H4,H5}     0000 0003 0000 0060     PAT{H4,H5}  EE
	{H6,H7}     0003 0000 0006 0000     PAT{H6,H7}  BB
	{H8,H9}     0000 0005 8000 0002     PAT{H8,H9}  E6
\end{verbatim}
\end{small}

\section{Formal Specification of the MAA in the REC Language}
\label{ANNEX-REC}

	This annex presents the specification of the MAA in the REC language. This specification is fully self-contained, meaning that it does not depend on any externally-defined library --- with the minor disadvantage of somewhat lengthy definitions for octet and blocks constants.

	For readability, the specification has been split into~21 parts, each part being devoted to a particular sort, a group of functions sharing a common purpose, or a collection of test vectors. The first parts contain general definitions that are largely independent from the MAA; starting from Sect.~\ref{ANNEX-REC-SPECIFIC}, the definitions become increasingly MAA-specific.

	The complete REC specification is obtained by concatenating all these parts, grouping their various sections (i.e., merging all \T{SORTS} sections into a single one, all \T{CONS} sections into a single one, etc.) and, after this, removing duplicated variable declarations in the \T{VARS} section, and pasting broken lines so that each rewrite rule that was split across several lines now fits on one single line.

	For readability, the rewrite rules concerning the same non-constructors have been put together and separated with blank lines when appropriate. Also, the arguments of certain non-constructors are separated by semicolons rather than commas when it helps to distinguish arguments of different nature (e.g., summand bits, carry, or sum bits).

	All machine words (octets, blocks, etc.) are represented according to the ``big endian'' convention, i.e., the first argument of each corresponding constructor denote the most significant bit.

\subsection{Definitions for sort Bool}

We first define Booleans using the \T{false} and \T{true} constructors, together with two non-constructors implementing logical conjunction and disjunction.

\begin{small}
\begin{verbatim}
 SORTS
   Bool
 CONS
   false : -> Bool
   true : -> Bool
 OPNS
   andBool : Bool Bool -> Bool
   orBool : Bool Bool -> Bool
 VARS
   L : Bool
 RULES
   andBool (false, L) -> false
   andBool (true, L) -> L

   orBool (false, L) -> L
   orBool (true, L) -> true
\end{verbatim}
\end{small}

\subsection{Definitions for sort Nat}

We then define natural numbers in the Peano style using the \T{zero} and \T{succ} constructors, together with non-constructors implementing addition, multiplication, equality, strict order, and a few constants.

\begin{small}
\begin{verbatim}
 SORTS
   Nat
 CONS
   zero : -> Nat
   succ : Nat -> Nat
 OPNS
   addNat : Nat Nat -> Nat
   multNat : Nat Nat -> Nat
   eqNat : Nat Nat -> Bool
   ltNat : Nat Nat -> Bool
   n1 : -> Nat
   n2 : -> Nat
   n3 : -> Nat
   n4 : -> Nat
   n5 : -> Nat
   n6 : -> Nat
   n7 : -> Nat
   n8 : -> Nat
   n9 : -> Nat
   n10 : -> Nat
   n11 : -> Nat
   n12 : -> Nat
   n13 : -> Nat
   n14 : -> Nat
   n15 : -> Nat
   n16 : -> Nat
   n17 : -> Nat
   n18 : -> Nat
   n19 : -> Nat
   n20 : -> Nat
   n21 : -> Nat
   n22 : -> Nat
   n254 : -> Nat
   n256 : -> Nat
   n4100 : -> Nat
 VARS
   N N' : Nat
 RULES
   addNat (N, zero) -> N
   addNat (N, succ (N')) -> addNat (succ (N), N')

   multNat (N, zero) -> zero
   multNat (N, succ (N')) -> addNat (N, multNat (N, N'))

   eqNat (zero, zero) -> true
   eqNat (zero, succ (N')) -> false
   eqNat (succ (N), zero) -> false
   eqNat (succ (N), succ (N')) -> eqNat (N, N')

   ltNat (zero, zero) -> false
   ltNat (zero, succ (N')) -> true
   ltNat (succ (N'), zero) -> false
   ltNat (succ (N), succ (N')) -> ltNat (N, N')

   n1 -> succ (zero)
   n2 -> succ (n1)
   n3 -> succ (n2)
   n4 -> succ (n3)
   n5 -> succ (n4)
   n6 -> succ (n5)
   n7 -> succ (n6)
   n8 -> succ (n7)
   n9 -> succ (n8)
   n10 -> succ (n9)
   n11 -> succ (n10)
   n12 -> succ (n11)
   n13 -> succ (n12)
   n14 -> succ (n13)
   n15 -> succ (n14)
   n16 -> succ (n15)
   n17 -> succ (n16)
   n18 -> succ (n17)
   n19 -> succ (n18)
   n20 -> succ (n19)
   n21 -> succ (n20)
   n22 -> succ (n21)

   n254 -> addNat (n12, multNat (n11, n22))

   n256 -> multNat (n16, n16)

   n4100 -> addNat (n4, multNat (n16, n256))
\end{verbatim}
\end{small}

\subsection{Definitions for sort Bit}

We now define bits using two constructors \T{x0} and \T{x1}, together with non-constructors implementing bit equality and logical operations on bits.

\begin{small}
\begin{verbatim}
 SORTS
   Bit
 CONS
   x0 : -> Bit
   x1 : -> Bit
 OPNS
   eqBit : Bit Bit -> Bool
   notBit : Bit -> Bit
   andBit : Bit Bit -> Bit
   orBit : Bit Bit -> Bit
   xorBit : Bit Bit -> Bit
 VARS
   B : Bit
 RULES
   eqBit (x0, x0) -> true
   eqBit (x0, x1) -> false
   eqBit (x1, x0) -> false
   eqBit (x1, x1) -> true

   notBit (x0) -> x1
   notBit (x1) -> x0

   andBit (B, x0) -> x0
   andBit (B, x1) -> B

   orBit (B, x0) -> B
   orBit (B, x1) -> x1

   xorBit (B, x0) -> B
   xorBit (B, x1) -> notBit (B)
\end{verbatim}
\end{small}

\subsection{Definitions for sort Octet}

We now define octets using a constructor \T{buildOctet} that takes eight bits and returns a byte, together with non-constructors implementing equality, bitwise logical operations, left-shift and right-shift operations on octets, as well as all octet constants needed to formally describe the MAA and its test vectors.

\begin{small}
\begin{verbatim}
 SORTS
   Octet
 CONS
   buildOctet : Bit Bit Bit Bit Bit Bit Bit Bit -> Octet
   % the first argument of buildOctet contains the most significant bit
 OPNS
   eqOctet : Octet Octet -> Bool
   andOctet : Octet Octet -> Octet
   orOctet : Octet Octet -> Octet
   xorOctet : Octet Octet -> Octet
   leftOctet1 : Octet -> Octet
   leftOctet2 : Octet -> Octet
   leftOctet3 : Octet -> Octet
   leftOctet4 : Octet -> Octet
   leftOctet5 : Octet -> Octet
   leftOctet6 : Octet -> Octet
   leftOctet7 : Octet -> Octet
   rightOctet1 : Octet -> Octet
   rightOctet2 : Octet -> Octet
   rightOctet3 : Octet -> Octet
   rightOctet4 : Octet -> Octet
   rightOctet5 : Octet -> Octet
   rightOctet6 : Octet -> Octet
   rightOctet7 : Octet -> Octet
   x00 : -> Octet
   x01 : -> Octet
   x02 : -> Octet
   x03 : -> Octet
   x04 : -> Octet
   x05 : -> Octet
   x06 : -> Octet
   x07 : -> Octet
   x08 : -> Octet
   x09 : -> Octet
   x0A : -> Octet
   x0B : -> Octet
   x0C : -> Octet
   x0D : -> Octet
   x0E : -> Octet
   x0F : -> Octet
   x10 : -> Octet
   x11 : -> Octet
   x12 : -> Octet
   x13 : -> Octet
   x14 : -> Octet
   x15 : -> Octet
   x16 : -> Octet
   x17 : -> Octet
   x18 : -> Octet
   x1A : -> Octet
   x1B : -> Octet
   x1C : -> Octet
   x1D : -> Octet
   x1E : -> Octet
   x1F : -> Octet
   x20 : -> Octet
   x21 : -> Octet
   x23 : -> Octet
   x24 : -> Octet
   x25 : -> Octet
   x26 : -> Octet
   x27 : -> Octet
   x28 : -> Octet
   x29 : -> Octet
   x2A : -> Octet
   x2B : -> Octet
   x2D : -> Octet
   x2E : -> Octet
   x2F : -> Octet
   x30 : -> Octet
   x31 : -> Octet
   x32 : -> Octet
   x33 : -> Octet
   x34 : -> Octet
   x35 : -> Octet
   x36 : -> Octet
   x37 : -> Octet
   x38 : -> Octet
   x39 : -> Octet
   x3A : -> Octet
   x3B : -> Octet
   x3C : -> Octet
   x3D : -> Octet
   x3F : -> Octet
   x40 : -> Octet
   x46 : -> Octet
   x48 : -> Octet
   x49 : -> Octet
   x4A : -> Octet
   x4B : -> Octet
   x4C : -> Octet
   x4D : -> Octet
   x4E : -> Octet
   x4F : -> Octet
   x50 : -> Octet
   x51 : -> Octet
   x53 : -> Octet
   x54 : -> Octet
   x55 : -> Octet
   x58 : -> Octet
   x5A : -> Octet
   x5B : -> Octet
   x5C : -> Octet
   x5D : -> Octet
   x5E : -> Octet
   x5F : -> Octet
   x60 : -> Octet
   x61 : -> Octet
   x62 : -> Octet
   x63 : -> Octet
   x64 : -> Octet
   x65 : -> Octet
   x66 : -> Octet
   x67 : -> Octet
   x69 : -> Octet
   x6A : -> Octet
   x6B : -> Octet
   x6C : -> Octet
   x6D : -> Octet
   x6E : -> Octet
   x6F : -> Octet
   x70 : -> Octet
   x71 : -> Octet
   x72 : -> Octet
   x73 : -> Octet
   x74 : -> Octet
   x75 : -> Octet
   x76 : -> Octet
   x77 : -> Octet
   x78 : -> Octet
   x79 : -> Octet
   x7A : -> Octet
   x7B : -> Octet
   x7C : -> Octet
   x7D : -> Octet
   x7E : -> Octet
   x7F : -> Octet
   x80 : -> Octet
   x81 : -> Octet
   x83 : -> Octet
   x84 : -> Octet
   x85 : -> Octet
   x86 : -> Octet
   x88 : -> Octet
   x89 : -> Octet
   x8A : -> Octet
   x8C : -> Octet
   x8D : -> Octet
   x8E : -> Octet
   x8F : -> Octet
   x90 : -> Octet
   x91 : -> Octet
   x92 : -> Octet
   x93 : -> Octet
   x95 : -> Octet
   x96 : -> Octet
   x97 : -> Octet
   x98 : -> Octet
   x99 : -> Octet
   x9A : -> Octet
   x9B : -> Octet
   x9C : -> Octet
   x9D : -> Octet
   x9E : -> Octet
   x9F : -> Octet
   xA0 : -> Octet
   xA1 : -> Octet
   xA2 : -> Octet
   xA3 : -> Octet
   xA4 : -> Octet
   xA5 : -> Octet
   xA6 : -> Octet
   xA7 : -> Octet
   xA8 : -> Octet
   xA9 : -> Octet
   xAA : -> Octet
   xAB : -> Octet
   xAC : -> Octet
   xAE : -> Octet
   xAF : -> Octet
   xB0 : -> Octet
   xB1 : -> Octet
   xB2 : -> Octet
   xB3 : -> Octet
   xB5 : -> Octet
   xB6 : -> Octet
   xB8 : -> Octet
   xB9 : -> Octet
   xBA : -> Octet
   xBB : -> Octet
   xBC : -> Octet
   xBE : -> Octet
   xBF : -> Octet
   xC0 : -> Octet
   xC1 : -> Octet
   xC2 : -> Octet
   xC4 : -> Octet
   xC5 : -> Octet
   xC6 : -> Octet
   xC7 : -> Octet
   xC8 : -> Octet
   xC9 : -> Octet
   xCA : -> Octet
   xCB : -> Octet
   xCC : -> Octet
   xCD : -> Octet
   xCE : -> Octet
   xD0 : -> Octet
   xD1 : -> Octet
   xD2 : -> Octet
   xD3 : -> Octet
   xD4 : -> Octet
   xD5 : -> Octet
   xD6 : -> Octet
   xD7 : -> Octet
   xD8 : -> Octet
   xD9 : -> Octet
   xDB : -> Octet
   xDC : -> Octet
   xDD : -> Octet
   xDE : -> Octet
   xDF : -> Octet
   xE0 : -> Octet
   xE1 : -> Octet
   xE3 : -> Octet
   xE6 : -> Octet
   xE8 : -> Octet
   xE9 : -> Octet
   xEA : -> Octet
   xEB : -> Octet
   xEC : -> Octet
   xED : -> Octet
   xEE : -> Octet
   xEF : -> Octet
   xF0 : -> Octet
   xF1 : -> Octet
   xF2 : -> Octet
   xF3 : -> Octet
   xF4 : -> Octet
   xF5 : -> Octet
   xF6 : -> Octet
   xF7 : -> Octet
   xF8 : -> Octet
   xF9 : -> Octet
   xFA : -> Octet
   xFB : -> Octet
   xFC : -> Octet
   xFD : -> Octet
   xFE : -> Octet
   xFF : -> Octet
 VARS
   B1 B2 B3 B4 B5 B6 B7 B8 : Bit
   B'1 B'2 B'3 B'4 B'5 B'6 B'7 B'8 : Bit
 RULES
   eqOctet (buildOctet (B1, B2, B3, B4, B5, B6, B7, B8),
            buildOctet (B'1, B'2, B'3, B'4, B'5, B'6, B'7, B'8))
   -> andBool (eqBit (B1, B'1), andBool (eqBit (B2, B'2),
      andBool (eqBit (B3, B'3), andBool (eqBit (B4, B'4),
      andBool (eqBit (B5, B'5), andBool (eqBit (B6, B'6),
      andBool (eqBit (B7, B'7), eqBit (B8, B'8))))))))

   andOctet (buildOctet (B1, B2, B3, B4, B5, B6, B7, B8),
             buildOctet (B'1, B'2, B'3, B'4, B'5, B'6, B'7, B'8))
   -> buildOctet (andBit (B1, B'1), andBit (B2, B'2),
                  andBit (B3, B'3), andBit (B4, B'4),
                  andBit (B5, B'5), andBit (B6, B'6),
                  andBit (B7, B'7), andBit (B8, B'8))

   orOctet (buildOctet (B1, B2, B3, B4, B5, B6, B7, B8),
            buildOctet (B'1, B'2, B'3, B'4, B'5, B'6, B'7, B'8))
   -> buildOctet (orBit (B1, B'1), orBit (B2, B'2),
                  orBit (B3, B'3), orBit (B4, B'4),
                  orBit (B5, B'5), orBit (B6, B'6),
                  orBit (B7, B'7), orBit (B8, B'8))

   xorOctet (buildOctet (B1, B2, B3, B4, B5, B6, B7, B8),
             buildOctet (B'1, B'2, B'3, B'4, B'5, B'6, B'7, B'8))
   -> buildOctet (xorBit (B1, B'1), xorBit (B2, B'2),
                  xorBit (B3, B'3), xorBit (B4, B'4),
                  xorBit (B5, B'5), xorBit (B6, B'6),
                  xorBit (B7, B'7), xorBit (B8, B'8))

   leftOctet1 (buildOctet (B1, B2, B3, B4, B5, B6, B7, B8))
   ->          buildOctet (B2, B3, B4, B5, B6, B7, B8, x0)

   leftOctet2 (buildOctet (B1, B2, B3, B4, B5, B6, B7, B8))
   ->          buildOctet (B3, B4, B5, B6, B7, B8, x0, x0)

   leftOctet3 (buildOctet (B1, B2, B3, B4, B5, B6, B7, B8))
   ->          buildOctet (B4, B5, B6, B7, B8, x0, x0, x0)

   leftOctet4 (buildOctet (B1, B2, B3, B4, B5, B6, B7, B8))
   ->          buildOctet (B5, B6, B7, B8, x0, x0, x0, x0)

   leftOctet5 (buildOctet (B1, B2, B3, B4, B5, B6, B7, B8))
   ->          buildOctet (B6, B7, B8, x0, x0, x0, x0, x0)

   leftOctet6 (buildOctet (B1, B2, B3, B4, B5, B6, B7, B8))
   ->          buildOctet (B7, B8, x0, x0, x0, x0, x0, x0)

   leftOctet7 (buildOctet (B1, B2, B3, B4, B5, B6, B7, B8))
   ->          buildOctet (B8, x0, x0, x0, x0, x0, x0, x0)

   rightOctet1 (buildOctet (B1, B2, B3, B4, B5, B6, B7, B8))
   ->           buildOctet (x0, B1, B2, B3, B4, B5, B6, B7)

   rightOctet2 (buildOctet (B1, B2, B3, B4, B5, B6, B7, B8))
   ->           buildOctet (x0, x0, B1, B2, B3, B4, B5, B6)

   rightOctet3 (buildOctet (B1, B2, B3, B4, B5, B6, B7, B8))
   ->           buildOctet (x0, x0, x0, B1, B2, B3, B4, B5)

   rightOctet4 (buildOctet (B1, B2, B3, B4, B5, B6, B7, B8))
   ->           buildOctet (x0, x0, x0, x0, B1, B2, B3, B4)

   rightOctet5 (buildOctet (B1, B2, B3, B4, B5, B6, B7, B8))
   ->           buildOctet (x0, x0, x0, x0, x0, B1, B2, B3)

   rightOctet6 (buildOctet (B1, B2, B3, B4, B5, B6, B7, B8))
   ->           buildOctet (x0, x0, x0, x0, x0, x0, B1, B2)

   rightOctet7 (buildOctet (B1, B2, B3, B4, B5, B6, B7, B8))
   ->           buildOctet (x0, x0, x0, x0, x0, x0, x0, B1)

   x00 -> buildOctet (x0, x0, x0, x0, x0, x0, x0, x0)
   x01 -> buildOctet (x0, x0, x0, x0, x0, x0, x0, x1)
   x02 -> buildOctet (x0, x0, x0, x0, x0, x0, x1, x0)
   x03 -> buildOctet (x0, x0, x0, x0, x0, x0, x1, x1)
   x04 -> buildOctet (x0, x0, x0, x0, x0, x1, x0, x0)
   x05 -> buildOctet (x0, x0, x0, x0, x0, x1, x0, x1)
   x06 -> buildOctet (x0, x0, x0, x0, x0, x1, x1, x0)
   x07 -> buildOctet (x0, x0, x0, x0, x0, x1, x1, x1)
   x08 -> buildOctet (x0, x0, x0, x0, x1, x0, x0, x0)
   x09 -> buildOctet (x0, x0, x0, x0, x1, x0, x0, x1)
   x0A -> buildOctet (x0, x0, x0, x0, x1, x0, x1, x0)
   x0B -> buildOctet (x0, x0, x0, x0, x1, x0, x1, x1)
   x0C -> buildOctet (x0, x0, x0, x0, x1, x1, x0, x0)
   x0D -> buildOctet (x0, x0, x0, x0, x1, x1, x0, x1)
   x0E -> buildOctet (x0, x0, x0, x0, x1, x1, x1, x0)
   x0F -> buildOctet (x0, x0, x0, x0, x1, x1, x1, x1)
   x10 -> buildOctet (x0, x0, x0, x1, x0, x0, x0, x0)
   x11 -> buildOctet (x0, x0, x0, x1, x0, x0, x0, x1)
   x12 -> buildOctet (x0, x0, x0, x1, x0, x0, x1, x0)
   x13 -> buildOctet (x0, x0, x0, x1, x0, x0, x1, x1)
   x14 -> buildOctet (x0, x0, x0, x1, x0, x1, x0, x0)
   x15 -> buildOctet (x0, x0, x0, x1, x0, x1, x0, x1)
   x16 -> buildOctet (x0, x0, x0, x1, x0, x1, x1, x0)
   x17 -> buildOctet (x0, x0, x0, x1, x0, x1, x1, x1)
   x18 -> buildOctet (x0, x0, x0, x1, x1, x0, x0, x0)
   x1A -> buildOctet (x0, x0, x0, x1, x1, x0, x1, x0)
   x1B -> buildOctet (x0, x0, x0, x1, x1, x0, x1, x1)
   x1C -> buildOctet (x0, x0, x0, x1, x1, x1, x0, x0)
   x1D -> buildOctet (x0, x0, x0, x1, x1, x1, x0, x1)
   x1E -> buildOctet (x0, x0, x0, x1, x1, x1, x1, x0)
   x1F -> buildOctet (x0, x0, x0, x1, x1, x1, x1, x1)
   x20 -> buildOctet (x0, x0, x1, x0, x0, x0, x0, x0)
   x21 -> buildOctet (x0, x0, x1, x0, x0, x0, x0, x1)
   x23 -> buildOctet (x0, x0, x1, x0, x0, x0, x1, x1)
   x24 -> buildOctet (x0, x0, x1, x0, x0, x1, x0, x0)
   x25 -> buildOctet (x0, x0, x1, x0, x0, x1, x0, x1)
   x26 -> buildOctet (x0, x0, x1, x0, x0, x1, x1, x0)
   x27 -> buildOctet (x0, x0, x1, x0, x0, x1, x1, x1)
   x28 -> buildOctet (x0, x0, x1, x0, x1, x0, x0, x0)
   x29 -> buildOctet (x0, x0, x1, x0, x1, x0, x0, x1)
   x2A -> buildOctet (x0, x0, x1, x0, x1, x0, x1, x0)
   x2B -> buildOctet (x0, x0, x1, x0, x1, x0, x1, x1)
   x2D -> buildOctet (x0, x0, x1, x0, x1, x1, x0, x1)
   x2E -> buildOctet (x0, x0, x1, x0, x1, x1, x1, x0)
   x2F -> buildOctet (x0, x0, x1, x0, x1, x1, x1, x1)
   x30 -> buildOctet (x0, x0, x1, x1, x0, x0, x0, x0)
   x31 -> buildOctet (x0, x0, x1, x1, x0, x0, x0, x1)
   x32 -> buildOctet (x0, x0, x1, x1, x0, x0, x1, x0)
   x33 -> buildOctet (x0, x0, x1, x1, x0, x0, x1, x1)
   x34 -> buildOctet (x0, x0, x1, x1, x0, x1, x0, x0)
   x35 -> buildOctet (x0, x0, x1, x1, x0, x1, x0, x1)
   x36 -> buildOctet (x0, x0, x1, x1, x0, x1, x1, x0)
   x37 -> buildOctet (x0, x0, x1, x1, x0, x1, x1, x1)
   x38 -> buildOctet (x0, x0, x1, x1, x1, x0, x0, x0)
   x39 -> buildOctet (x0, x0, x1, x1, x1, x0, x0, x1)
   x3A -> buildOctet (x0, x0, x1, x1, x1, x0, x1, x0)
   x3B -> buildOctet (x0, x0, x1, x1, x1, x0, x1, x1)
   x3C -> buildOctet (x0, x0, x1, x1, x1, x1, x0, x0)
   x3D -> buildOctet (x0, x0, x1, x1, x1, x1, x0, x1)
   x3F -> buildOctet (x0, x0, x1, x1, x1, x1, x1, x1)
   x40 -> buildOctet (x0, x1, x0, x0, x0, x0, x0, x0)
   x46 -> buildOctet (x0, x1, x0, x0, x0, x1, x1, x0)
   x48 -> buildOctet (x0, x1, x0, x0, x1, x0, x0, x0)
   x49 -> buildOctet (x0, x1, x0, x0, x1, x0, x0, x1)
   x4A -> buildOctet (x0, x1, x0, x0, x1, x0, x1, x0)
   x4B -> buildOctet (x0, x1, x0, x0, x1, x0, x1, x1)
   x4C -> buildOctet (x0, x1, x0, x0, x1, x1, x0, x0)
   x4D -> buildOctet (x0, x1, x0, x0, x1, x1, x0, x1)
   x4E -> buildOctet (x0, x1, x0, x0, x1, x1, x1, x0)
   x4F -> buildOctet (x0, x1, x0, x0, x1, x1, x1, x1)
   x50 -> buildOctet (x0, x1, x0, x1, x0, x0, x0, x0)
   x51 -> buildOctet (x0, x1, x0, x1, x0, x0, x0, x1)
   x53 -> buildOctet (x0, x1, x0, x1, x0, x0, x1, x1)
   x54 -> buildOctet (x0, x1, x0, x1, x0, x1, x0, x0)
   x55 -> buildOctet (x0, x1, x0, x1, x0, x1, x0, x1)
   x58 -> buildOctet (x0, x1, x0, x1, x1, x0, x0, x0)
   x5A -> buildOctet (x0, x1, x0, x1, x1, x0, x1, x0)
   x5B -> buildOctet (x0, x1, x0, x1, x1, x0, x1, x1)
   x5C -> buildOctet (x0, x1, x0, x1, x1, x1, x0, x0)
   x5D -> buildOctet (x0, x1, x0, x1, x1, x1, x0, x1)
   x5E -> buildOctet (x0, x1, x0, x1, x1, x1, x1, x0)
   x5F -> buildOctet (x0, x1, x0, x1, x1, x1, x1, x1)
   x60 -> buildOctet (x0, x1, x1, x0, x0, x0, x0, x0)
   x61 -> buildOctet (x0, x1, x1, x0, x0, x0, x0, x1)
   x62 -> buildOctet (x0, x1, x1, x0, x0, x0, x1, x0)
   x63 -> buildOctet (x0, x1, x1, x0, x0, x0, x1, x1)
   x64 -> buildOctet (x0, x1, x1, x0, x0, x1, x0, x0)
   x65 -> buildOctet (x0, x1, x1, x0, x0, x1, x0, x1)
   x66 -> buildOctet (x0, x1, x1, x0, x0, x1, x1, x0)
   x67 -> buildOctet (x0, x1, x1, x0, x0, x1, x1, x1)
   x69 -> buildOctet (x0, x1, x1, x0, x1, x0, x0, x1)
   x6A -> buildOctet (x0, x1, x1, x0, x1, x0, x1, x0)
   x6B -> buildOctet (x0, x1, x1, x0, x1, x0, x1, x1)
   x6C -> buildOctet (x0, x1, x1, x0, x1, x1, x0, x0)
   x6D -> buildOctet (x0, x1, x1, x0, x1, x1, x0, x1)
   x6E -> buildOctet (x0, x1, x1, x0, x1, x1, x1, x0)
   x6F -> buildOctet (x0, x1, x1, x0, x1, x1, x1, x1)
   x70 -> buildOctet (x0, x1, x1, x1, x0, x0, x0, x0)
   x71 -> buildOctet (x0, x1, x1, x1, x0, x0, x0, x1)
   x72 -> buildOctet (x0, x1, x1, x1, x0, x0, x1, x0)
   x73 -> buildOctet (x0, x1, x1, x1, x0, x0, x1, x1)
   x74 -> buildOctet (x0, x1, x1, x1, x0, x1, x0, x0)
   x75 -> buildOctet (x0, x1, x1, x1, x0, x1, x0, x1)
   x76 -> buildOctet (x0, x1, x1, x1, x0, x1, x1, x0)
   x77 -> buildOctet (x0, x1, x1, x1, x0, x1, x1, x1)
   x78 -> buildOctet (x0, x1, x1, x1, x1, x0, x0, x0)
   x79 -> buildOctet (x0, x1, x1, x1, x1, x0, x0, x1)
   x7A -> buildOctet (x0, x1, x1, x1, x1, x0, x1, x0)
   x7B -> buildOctet (x0, x1, x1, x1, x1, x0, x1, x1)
   x7C -> buildOctet (x0, x1, x1, x1, x1, x1, x0, x0)
   x7D -> buildOctet (x0, x1, x1, x1, x1, x1, x0, x1)
   x7E -> buildOctet (x0, x1, x1, x1, x1, x1, x1, x0)
   x7F -> buildOctet (x0, x1, x1, x1, x1, x1, x1, x1)
   x80 -> buildOctet (x1, x0, x0, x0, x0, x0, x0, x0)
   x81 -> buildOctet (x1, x0, x0, x0, x0, x0, x0, x1)
   x83 -> buildOctet (x1, x0, x0, x0, x0, x0, x1, x1)
   x84 -> buildOctet (x1, x0, x0, x0, x0, x1, x0, x0)
   x85 -> buildOctet (x1, x0, x0, x0, x0, x1, x0, x1)
   x86 -> buildOctet (x1, x0, x0, x0, x0, x1, x1, x0)
   x88 -> buildOctet (x1, x0, x0, x0, x1, x0, x0, x0)
   x89 -> buildOctet (x1, x0, x0, x0, x1, x0, x0, x1)
   x8A -> buildOctet (x1, x0, x0, x0, x1, x0, x1, x0)
   x8C -> buildOctet (x1, x0, x0, x0, x1, x1, x0, x0)
   x8D -> buildOctet (x1, x0, x0, x0, x1, x1, x0, x1)
   x8E -> buildOctet (x1, x0, x0, x0, x1, x1, x1, x0)
   x8F -> buildOctet (x1, x0, x0, x0, x1, x1, x1, x1)
   x90 -> buildOctet (x1, x0, x0, x1, x0, x0, x0, x0)
   x91 -> buildOctet (x1, x0, x0, x1, x0, x0, x0, x1)
   x92 -> buildOctet (x1, x0, x0, x1, x0, x0, x1, x0)
   x93 -> buildOctet (x1, x0, x0, x1, x0, x0, x1, x1)
   x95 -> buildOctet (x1, x0, x0, x1, x0, x1, x0, x1)
   x96 -> buildOctet (x1, x0, x0, x1, x0, x1, x1, x0)
   x97 -> buildOctet (x1, x0, x0, x1, x0, x1, x1, x1)
   x98 -> buildOctet (x1, x0, x0, x1, x1, x0, x0, x0)
   x99 -> buildOctet (x1, x0, x0, x1, x1, x0, x0, x1)
   x9A -> buildOctet (x1, x0, x0, x1, x1, x0, x1, x0)
   x9B -> buildOctet (x1, x0, x0, x1, x1, x0, x1, x1)
   x9C -> buildOctet (x1, x0, x0, x1, x1, x1, x0, x0)
   x9D -> buildOctet (x1, x0, x0, x1, x1, x1, x0, x1)
   x9E -> buildOctet (x1, x0, x0, x1, x1, x1, x1, x0)
   x9F -> buildOctet (x1, x0, x0, x1, x1, x1, x1, x1)
   xA0 -> buildOctet (x1, x0, x1, x0, x0, x0, x0, x0)
   xA1 -> buildOctet (x1, x0, x1, x0, x0, x0, x0, x1)
   xA2 -> buildOctet (x1, x0, x1, x0, x0, x0, x1, x0)
   xA3 -> buildOctet (x1, x0, x1, x0, x0, x0, x1, x1)
   xA4 -> buildOctet (x1, x0, x1, x0, x0, x1, x0, x0)
   xA5 -> buildOctet (x1, x0, x1, x0, x0, x1, x0, x1)
   xA6 -> buildOctet (x1, x0, x1, x0, x0, x1, x1, x0)
   xA7 -> buildOctet (x1, x0, x1, x0, x0, x1, x1, x1)
   xA8 -> buildOctet (x1, x0, x1, x0, x1, x0, x0, x0)
   xA9 -> buildOctet (x1, x0, x1, x0, x1, x0, x0, x1)
   xAA -> buildOctet (x1, x0, x1, x0, x1, x0, x1, x0)
   xAB -> buildOctet (x1, x0, x1, x0, x1, x0, x1, x1)
   xAC -> buildOctet (x1, x0, x1, x0, x1, x1, x0, x0)
   xAE -> buildOctet (x1, x0, x1, x0, x1, x1, x1, x0)
   xAF -> buildOctet (x1, x0, x1, x0, x1, x1, x1, x1)
   xB0 -> buildOctet (x1, x0, x1, x1, x0, x0, x0, x0)
   xB1 -> buildOctet (x1, x0, x1, x1, x0, x0, x0, x1)
   xB2 -> buildOctet (x1, x0, x1, x1, x0, x0, x1, x0)
   xB3 -> buildOctet (x1, x0, x1, x1, x0, x0, x1, x1)
   xB5 -> buildOctet (x1, x0, x1, x1, x0, x1, x0, x1)
   xB6 -> buildOctet (x1, x0, x1, x1, x0, x1, x1, x0)
   xB8 -> buildOctet (x1, x0, x1, x1, x1, x0, x0, x0)
   xB9 -> buildOctet (x1, x0, x1, x1, x1, x0, x0, x1)
   xBA -> buildOctet (x1, x0, x1, x1, x1, x0, x1, x0)
   xBB -> buildOctet (x1, x0, x1, x1, x1, x0, x1, x1)
   xBC -> buildOctet (x1, x0, x1, x1, x1, x1, x0, x0)
   xBE -> buildOctet (x1, x0, x1, x1, x1, x1, x1, x0)
   xBF -> buildOctet (x1, x0, x1, x1, x1, x1, x1, x1)
   xC0 -> buildOctet (x1, x1, x0, x0, x0, x0, x0, x0)
   xC1 -> buildOctet (x1, x1, x0, x0, x0, x0, x0, x1)
   xC2 -> buildOctet (x1, x1, x0, x0, x0, x0, x1, x0)
   xC4 -> buildOctet (x1, x1, x0, x0, x0, x1, x0, x0)
   xC5 -> buildOctet (x1, x1, x0, x0, x0, x1, x0, x1)
   xC6 -> buildOctet (x1, x1, x0, x0, x0, x1, x1, x0)
   xC7 -> buildOctet (x1, x1, x0, x0, x0, x1, x1, x1)
   xC8 -> buildOctet (x1, x1, x0, x0, x1, x0, x0, x0)
   xC9 -> buildOctet (x1, x1, x0, x0, x1, x0, x0, x1)
   xCA -> buildOctet (x1, x1, x0, x0, x1, x0, x1, x0)
   xCB -> buildOctet (x1, x1, x0, x0, x1, x0, x1, x1)
   xCC -> buildOctet (x1, x1, x0, x0, x1, x1, x0, x0)
   xCD -> buildOctet (x1, x1, x0, x0, x1, x1, x0, x1)
   xCE -> buildOctet (x1, x1, x0, x0, x1, x1, x1, x0)
   xD0 -> buildOctet (x1, x1, x0, x1, x0, x0, x0, x0)
   xD1 -> buildOctet (x1, x1, x0, x1, x0, x0, x0, x1)
   xD2 -> buildOctet (x1, x1, x0, x1, x0, x0, x1, x0)
   xD3 -> buildOctet (x1, x1, x0, x1, x0, x0, x1, x1)
   xD4 -> buildOctet (x1, x1, x0, x1, x0, x1, x0, x0)
   xD5 -> buildOctet (x1, x1, x0, x1, x0, x1, x0, x1)
   xD6 -> buildOctet (x1, x1, x0, x1, x0, x1, x1, x0)
   xD7 -> buildOctet (x1, x1, x0, x1, x0, x1, x1, x1)
   xD8 -> buildOctet (x1, x1, x0, x1, x1, x0, x0, x0)
   xD9 -> buildOctet (x1, x1, x0, x1, x1, x0, x0, x1)
   xDB -> buildOctet (x1, x1, x0, x1, x1, x0, x1, x1)
   xDC -> buildOctet (x1, x1, x0, x1, x1, x1, x0, x0)
   xDD -> buildOctet (x1, x1, x0, x1, x1, x1, x0, x1)
   xDE -> buildOctet (x1, x1, x0, x1, x1, x1, x1, x0)
   xDF -> buildOctet (x1, x1, x0, x1, x1, x1, x1, x1)
   xE0 -> buildOctet (x1, x1, x1, x0, x0, x0, x0, x0)
   xE1 -> buildOctet (x1, x1, x1, x0, x0, x0, x0, x1)
   xE3 -> buildOctet (x1, x1, x1, x0, x0, x0, x1, x1)
   xE6 -> buildOctet (x1, x1, x1, x0, x0, x1, x1, x0)
   xE8 -> buildOctet (x1, x1, x1, x0, x1, x0, x0, x0)
   xE9 -> buildOctet (x1, x1, x1, x0, x1, x0, x0, x1)
   xEA -> buildOctet (x1, x1, x1, x0, x1, x0, x1, x0)
   xEB -> buildOctet (x1, x1, x1, x0, x1, x0, x1, x1)
   xEC -> buildOctet (x1, x1, x1, x0, x1, x1, x0, x0)
   xED -> buildOctet (x1, x1, x1, x0, x1, x1, x0, x1)
   xEE -> buildOctet (x1, x1, x1, x0, x1, x1, x1, x0)
   xEF -> buildOctet (x1, x1, x1, x0, x1, x1, x1, x1)
   xF0 -> buildOctet (x1, x1, x1, x1, x0, x0, x0, x0)
   xF1 -> buildOctet (x1, x1, x1, x1, x0, x0, x0, x1)
   xF2 -> buildOctet (x1, x1, x1, x1, x0, x0, x1, x0)
   xF3 -> buildOctet (x1, x1, x1, x1, x0, x0, x1, x1)
   xF4 -> buildOctet (x1, x1, x1, x1, x0, x1, x0, x0)
   xF5 -> buildOctet (x1, x1, x1, x1, x0, x1, x0, x1)
   xF6 -> buildOctet (x1, x1, x1, x1, x0, x1, x1, x0)
   xF7 -> buildOctet (x1, x1, x1, x1, x0, x1, x1, x1)
   xF8 -> buildOctet (x1, x1, x1, x1, x1, x0, x0, x0)
   xF9 -> buildOctet (x1, x1, x1, x1, x1, x0, x0, x1)
   xFA -> buildOctet (x1, x1, x1, x1, x1, x0, x1, x0)
   xFB -> buildOctet (x1, x1, x1, x1, x1, x0, x1, x1)
   xFC -> buildOctet (x1, x1, x1, x1, x1, x1, x0, x0)
   xFD -> buildOctet (x1, x1, x1, x1, x1, x1, x0, x1)
   xFE -> buildOctet (x1, x1, x1, x1, x1, x1, x1, x0)
   xFF -> buildOctet (x1, x1, x1, x1, x1, x1, x1, x1)
\end{verbatim}
\end{small}

\subsection{Definitions for sort OctetSum}

We now define sort \T{OctetSum} that stores the result of the addition of two octets. Values of this sort are 9-bit words, made up using the constructor \T{buildOctetSum} that gathers one bit for the carry and an octet for the sum. The three principal non-constructors for this sort are \T{eqOctetSum} (which tests equality), \T{addOctetSum} (which adds two octets and an input carry bit, and returns both an output carry bit and an 8-bit sum), and \T{addOctet} (which is derived from the former one by dropping the input and output carry bits); the other non-constructors are auxiliary functions implementing an 8-bit adder.

\begin{small}
\begin{verbatim}
 SORTS
   OctetSum
 CONS
   buildOctetSum : Bit Octet -> OctetSum
 OPNS
   eqOctetSum : OctetSum OctetSum -> Bool
   addBit : Bit Bit Bit -> Bit
   carBit : Bit Bit Bit -> Bit
   addOctetSum : Octet Octet Bit -> OctetSum
   addOctet8 : Bit Bit Bit Bit Bit Bit Bit Bit Bit Bit Bit Bit Bit Bit Bit Bit Bit
               -> OctetSum
   addOctet7 : Bit Bit Bit Bit Bit Bit Bit Bit Bit Bit Bit Bit Bit Bit Bit Bit
               -> OctetSum
   addOctet6 : Bit Bit Bit Bit Bit Bit Bit Bit Bit Bit Bit Bit Bit Bit Bit -> OctetSum
   addOctet5 : Bit Bit Bit Bit Bit Bit Bit Bit Bit Bit Bit Bit Bit Bit -> OctetSum
   addOctet4 : Bit Bit Bit Bit Bit Bit Bit Bit Bit Bit Bit Bit Bit -> OctetSum
   addOctet3 : Bit Bit Bit Bit Bit Bit Bit Bit Bit Bit Bit Bit -> OctetSum
   addOctet2 : Bit Bit Bit Bit Bit Bit Bit Bit Bit Bit Bit -> OctetSum
   addOctet1 : Bit Bit Bit Bit Bit Bit Bit Bit Bit Bit -> OctetSum
   addOctet0 : Bit Bit Bit Bit Bit Bit Bit Bit Bit -> OctetSum
   dropCarryOctetSum : OctetSum -> Octet
   addOctet : Octet Octet -> Octet
 VARS
   B B' Bcarry : Bit
   B1 B2 B3 B4 B5 B6 B7 B8 : Bit
   B'1 B'2 B'3 B'4 B'5 B'6 B'7 B'8 : Bit
   B"1 B"2 B"3 B"4 B"5 B"6 B"7 B"8 : Bit
   O O' : Octet
 RULES
   eqOctetSum (buildOctetSum (B, O), buildOctetSum (B', O'))
   -> andBool (eqBit (B, B'), eqOctet (O, O'))

   % addBit (B, B', Bcarry) is the sum of bits (B + B' + Bcarry) without carry
   addBit (B, B', Bcarry) -> xorBit (xorBit (B, B'), Bcarry)

   % carBit (B, B', Bcarry) is the carry for the sum of bits (B + B' + Bcarry)
   carBit (B, B', Bcarry) -> orBit (andBit (andBit (B, B'), notBit (Bcarry)),
                                    andBit (orBit (B, B'), Bcarry))

   addOctetSum (buildOctet (B1, B2, B3, B4, B5, B6, B7, B8),
                buildOctet (B'1, B'2, B'3, B'4, B'5, B'6, B'7, B'8), Bcarry)
   -> addOctet8 (B1, B'1, B2, B'2, B3, B'3, B4, B'4, B5, B'5, B6, B'6, B7, B'7, B8,
                 B'8; Bcarry)

   addOctet8 (B1, B'1, B2, B'2, B3, B'3, B4, B'4, B5, B'5, B6, B'6, B7, B'7, B8, B'8;
              Bcarry)
   -> addOctet7 (B1, B'1, B2, B'2, B3, B'3, B4, B'4, B5, B'5, B6, B'6, B7, B'7;
                 carBit (B8, B'8, Bcarry); addBit (B8, B'8, Bcarry))

   addOctet7 (B1, B'1, B2, B'2, B3, B'3, B4, B'4, B5, B'5, B6, B'6, B7, B'7;
              Bcarry; B"8)
   -> addOctet6 (B1, B'1, B2, B'2, B3, B'3, B4, B'4, B5, B'5, B6, B'6;
                 carBit (B7, B'7, Bcarry); addBit (B7, B'7, Bcarry), B"8)

   addOctet6 (B1, B'1, B2, B'2, B3, B'3, B4, B'4, B5, B'5, B6, B'6;
              Bcarry; B"7, B"8)
   -> addOctet5 (B1, B'1, B2, B'2, B3, B'3, B4, B'4, B5, B'5;
                 carBit (B6, B'6, Bcarry); addBit (B6, B'6, Bcarry), B"7, B"8)

   addOctet5 (B1, B'1, B2, B'2, B3, B'3, B4, B'4, B5, B'5;
              Bcarry; B"6, B"7, B"8)
   -> addOctet4 (B1, B'1, B2, B'2, B3, B'3, B4, B'4;
                 carBit (B5, B'5, Bcarry); addBit (B5, B'5, Bcarry), B"6, B"7, B"8)

   addOctet4 (B1, B'1, B2, B'2, B3, B'3, B4, B'4;
              Bcarry; B"5, B"6, B"7, B"8)
   -> addOctet3 (B1, B'1, B2, B'2, B3, B'3; carBit (B4, B'4, Bcarry);
                 addBit (B4, B'4, Bcarry), B"5, B"6, B"7, B"8)

   addOctet3 (B1, B'1, B2, B'2, B3, B'3;
              Bcarry; B"4, B"5, B"6, B"7, B"8)
   -> addOctet2 (B1, B'1, B2, B'2; carBit (B3, B'3, Bcarry);
                 addBit (B3, B'3, Bcarry), B"4, B"5, B"6, B"7, B"8)

   addOctet2 (B1, B'1, B2, B'2;
              Bcarry; B"3, B"4, B"5, B"6, B"7, B"8)
   -> addOctet1 (B1, B'1; carBit (B2, B'2, Bcarry);
                 addBit (B2, B'2, Bcarry), B"3, B"4, B"5, B"6, B"7, B"8)

   addOctet1 (B1, B'1;
              Bcarry; B"2, B"3, B"4, B"5, B"6, B"7, B"8)
   -> addOctet0 (carBit (B1, B'1, Bcarry);
                 addBit (B1, B'1, Bcarry), B"2, B"3, B"4, B"5, B"6, B"7, B"8)

   addOctet0 (Bcarry; B"1, B"2, B"3, B"4, B"5, B"6, B"7, B"8)
   -> buildOctetSum (Bcarry, buildOctet (B"1, B"2, B"3, B"4, B"5, B"6, B"7, B"8))

   dropCarryOctetSum (buildOctetSum (Bcarry, O)) -> O

   addOctet (O, O') -> dropCarryOctetSum (addOctetSum (O, O', x0))
\end{verbatim}
\end{small}

\subsection{Definitions for sort Half}

We now define 16-bit words (named ``half words'') using a constructor \T{buildHalf} that takes two octets and returns a half word, together with non-constructors implementing equality, two usual constants, and an operation \T{mulOctet} that takes two octets and computes their 16-bit product; the other non-constructors are auxiliary functions implementing an 8-bit multiplier.

\begin{small}
\begin{verbatim}
 SORTS
   Half
 CONS
   buildHalf : Octet Octet -> Half
   % the first argument of buildHalf contain the most significant bits
 OPNS
   eqHalf : Half Half -> Bool
   x0000 : -> Half
   x0001 : -> Half
   mulOctet : Octet Octet -> Half
   mulOctet1 : Bit Bit Bit Bit Bit Bit Bit Bit Octet Half -> Half
   mulOctet2 : Bit Bit Bit Bit Bit Bit Bit Octet Half -> Half
   mulOctet3 : Bit Bit Bit Bit Bit Bit Octet Half -> Half
   mulOctet4 : Bit Bit Bit Bit Bit Octet Half -> Half
   mulOctet5 : Bit Bit Bit Bit Octet Half -> Half
   mulOctet6 : Bit Bit Bit Octet Half -> Half
   mulOctet7 : Bit Bit Octet Half -> Half
   mulOctet8 : Bit Octet Half -> Half
   mulOctetA : Half Octet Octet -> Half
   mulOctetB : Octet OctetSum -> Half
 VARS
   B1 B2 B3 B4 B5 B6 B7 B8 : Bit
   O' O1 O2 O'1 O'2 : Octet
 RULES
   eqHalf (buildHalf (O1, O2), buildHalf (O'1, O'2))
   -> andBool (eqOctet (O1, O'1), eqOctet (O2, O'2))

   x0000 -> buildHalf (x00, x00)
   x0001 -> buildHalf (x00, x01)

   mulOctet (buildOctet (B1, B2, B3, B4, B5, B6, B7, B8), O')
   -> mulOctet1 (B1, B2, B3, B4, B5, B6, B7, B8, O', x0000)

   mulOctet1 (x0, B2, B3, B4, B5, B6, B7, B8, O', H) -> mulOctet2 (B2, B3, B4, B5, B6,
      B7, B8, O', H)
   mulOctet1 (x1, B2, B3, B4, B5, B6, B7, B8, O', H) -> mulOctet2 (B2, B3, B4, B5, B6,
      B7, B8, O', mulOctetA (H, rightOctet1 (O'), leftOctet7 (O')))

   mulOctet2 (x0, B3, B4, B5, B6, B7, B8, O', H) -> mulOctet3 (B3, B4, B5, B6, B7, B8,
      O', H)
   mulOctet2 (x1, B3, B4, B5, B6, B7, B8, O', H) -> mulOctet3 (B3, B4, B5, B6, B7, B8,
      O', mulOctetA (H, rightOctet2 (O'), leftOctet6 (O')))

   mulOctet3 (x0, B4, B5, B6, B7, B8, O', H) -> mulOctet4 (B4, B5, B6, B7, B8, O', H)
   mulOctet3 (x1, B4, B5, B6, B7, B8, O', H) -> mulOctet4 (B4, B5, B6, B7, B8, O',
      mulOctetA (H, rightOctet3 (O'), leftOctet5 (O')))

   mulOctet4 (x0, B5, B6, B7, B8, O', H) -> mulOctet5 (B5, B6, B7, B8, O', H)
   mulOctet4 (x1, B5, B6, B7, B8, O', H) -> mulOctet5 (B5, B6, B7, B8, O',
      mulOctetA (H, rightOctet4 (O'), leftOctet4 (O')))

   mulOctet5 (x0, B6, B7, B8, O', H) -> mulOctet6 (B6, B7, B8, O', H)
   mulOctet5 (x1, B6, B7, B8, O', H) -> mulOctet6 (B6, B7, B8, O',
      mulOctetA (H, rightOctet5 (O'), leftOctet3 (O')))

   mulOctet6 (x0, B7, B8, O', H) -> mulOctet7 (B7, B8, O', H)
   mulOctet6 (x1, B7, B8, O', H) -> mulOctet7 (B7, B8, O',
      mulOctetA (H, rightOctet6 (O'), leftOctet2 (O')))

   mulOctet7 (x0, B8, O', H) -> mulOctet8 (B8, O', H)
   mulOctet7 (x1, B8, O', H) -> mulOctet8 (B8, O',
      mulOctetA (H, rightOctet7 (O'), leftOctet1 (O')))

   mulOctet8 (x0, O', H) -> H
   mulOctet8 (x1, O', H) -> mulOctetA (H, x00, O')

   mulOctetA (buildHalf (O1, O2), O'1, O'2)
   -> mulOctetB (addOctet (O1, O'1), addOctetSum (O2, O'2, x0))

   mulOctetB (O1, buildOctetSum (x0, O2)) -> buildHalf (O1, O2)
   mulOctetB (O1, buildOctetSum (x1, O2)) -> buildHalf (addOctet (O1, x01), O2)
\end{verbatim}
\end{small}

\subsection{Definitions for sort HalfSum}

We now define sort \T{HalfSum} that stores the result of the addition of two half words. Values of this sort are 17-bit words, made up using the constructor \T{buildHalfSum} that gathers one bit for the carry and a half word for the sum. The five principal non-constructors for this sort are \T{eqHalfSum} (which tests equality), \T{addHalfSum} (which adds two half words and returns both a carry bit and a 16-bit sum), \T{addHalf} (which is derived from the former one by dropping the carry bit), \T{addHalfOctet} and \T{addHalfOctets} (which are similar to the former one but take octet arguments that are converted to half words before summation); the other non-constructors are auxiliary functions implementing a 16-bit adder built using two 8-bit adders.

\begin{small}
\begin{verbatim}
 SORTS
   HalfSum
 CONS
   buildHalfSum : Bit Half -> HalfSum
 OPNS
   eqHalfSum : HalfSum HalfSum -> Bool
   addHalfSum : Half Half -> HalfSum
   addHalf2 : Octet Octet Octet Octet -> HalfSum
   addHalf1 : Octet Octet OctetSum -> HalfSum
   addHalf0 : OctetSum Octet -> HalfSum
   dropCarryHalfSum : HalfSum -> Half
   addHalf : Half Half -> Half
   addHalfOctet : Octet Half -> Half
   addHalfOctets : Octet Octet -> Half
 VARS
   B B' : Bit
   O O' O1 O2 O'1 O'2 O"1 O"2 : Octet
   H H' : Half
 RULES
   eqHalfSum (buildHalfSum (B, H), buildHalfSum (B', H'))
   -> andBool (eqBit (B, B'), eqHalf (H, H'))

   addHalfSum (buildHalf (O1, O2), buildHalf (O'1, O'2)) -> addHalf2 (O1, O'1, O2, O'2)

   addHalf2 (O1, O'1, O2, O'2) -> addHalf1 (O1, O'1, addOctetSum (O2, O'2, x0))

   addHalf1 (O1, O'1, buildOctetSum (B,O"2)) -> addHalf0 (addOctetSum (O1, O'1, B),O"2)

   addHalf0 (buildOctetSum (B, O"1), O"2) -> buildHalfSum (B, buildHalf (O"1, O"2))

   dropCarryHalfSum (buildHalfSum (B, H)) -> H

   addHalf (H, H') -> dropCarryHalfSum (addHalfSum (H, H'))

   addHalfOctet (O, H) -> addHalf (buildHalf (x00, O), H)

   addHalfOctets (O, O') -> addHalf (buildHalf (x00, O), buildHalf (x00, O'))
\end{verbatim}
\end{small}

\subsection{Definitions for sort Block}

We now define 32-bit words (named ``blocks'' according to the MAA terminology) using a constructor \T{buildBlock} that takes four octets and returns a block. The seven principal non-constructors for this sort are \T{eqBlock} (which tests equality), \T{andBlock}, \T{orBlock}, and \T{xorBlock} (which implement bitwise logical operations on blocks), \T{HalfU} and \T{HalfL} (which decompose a block into two half words), and \T{mulHalf} (which takes two half words and computes their 32-bit product); the other non-constructors are auxiliary functions implementing a 16-bit multiplier built using four 8-bit multipliers, as well as all block constants needed to formally describe the MAA and its test vectors.

\begin{small}
\begin{verbatim}
 SORTS
   Block
 CONS
   buildBlock : Octet Octet Octet Octet -> Block
   % the first argument of buildBlock contain the most significant bits
 OPNS
   eqBlock : Block Block -> Bool
   andBlock : Block Block -> Block
   orBlock : Block Block -> Block
   xorBlock : Block Block -> Block
   HalfU : Block -> Half
   HalfL : Block -> Half
   mulHalf : Half Half -> Block
   mulHalfA : Half Half Half Half -> Block
   mulHalf4 : Octet Octet Octet Octet Octet Octet Octet Octet -> Block
   mulHalf3 : Octet Octet Octet Octet Half Octet -> Block
   mulHalf2 : Octet Half Octet Octet -> Block
   mulHalf1 : Half Octet Octet Octet -> Block
   x00000000 : -> Block
   x00000001 : -> Block
   x00000002 : -> Block
   x00000003 : -> Block
   x00000004 : -> Block
   x00000005 : -> Block
   x00000006 : -> Block
   x00000007 : -> Block
   x00000008 : -> Block
   x00000009 : -> Block
   x0000000A : -> Block
   x0000000B : -> Block
   x0000000C : -> Block
   x0000000D : -> Block
   x0000000E : -> Block
   x0000000F : -> Block
   x00000010 : -> Block
   x00000012 : -> Block
   x00000014 : -> Block
   x00000016 : -> Block
   x00000018 : -> Block
   x0000001B : -> Block
   x0000001D : -> Block
   x0000001E : -> Block
   x0000001F : -> Block
   x00000031 : -> Block
   x00000036 : -> Block
   x00000060 : -> Block
   x00000080 : -> Block
   x000000A5 : -> Block
   x000000B6 : -> Block
   x000000C4 : -> Block
   x000000D2 : -> Block
   x00000100 : -> Block
   x00000129 : -> Block
   x0000018C : -> Block
   x00004000 : -> Block
   x00010000 : -> Block
   x00020000 : -> Block
   x00030000 : -> Block
   x00040000 : -> Block
   x00060000 : -> Block
   x00804021 : -> Block   % MAA special constant 'B'
   x00FF00FF : -> Block
   x0103050B : -> Block
   x01030703 : -> Block
   x01030705 : -> Block
   x0103070F : -> Block
   x02040801 : -> Block   % MAA special constant 'A'
   x0297AF6F : -> Block
   x07050301 : -> Block
   x077788A2 : -> Block
   x07C72EAA : -> Block
   x0A202020 : -> Block
   x0AD67E20 : -> Block
   x10000000 : -> Block
   x11A9D254 : -> Block
   x11AC46B8 : -> Block
   x1277A6D4 : -> Block
   x13647149 : -> Block
   x160EE9B5 : -> Block
   x17065DBB : -> Block
   x17A808FD : -> Block
   x1D10D8D3 : -> Block
   x1D3B7760 : -> Block
   x1D9C9655 : -> Block
   x1F3F7FFF : -> Block
   x204E80A7 : -> Block
   x21D869BA : -> Block
   x24B66FB5 : -> Block
   x270EEDAF : -> Block
   x277B4B25 : -> Block
   x2829040B : -> Block
   x288FC786 : -> Block
   x28EAD8B3 : -> Block
   x29907CD8 : -> Block
   x29C1485F : -> Block
   x29EEE96B : -> Block
   x2A6091AE : -> Block
   x2BF8499A : -> Block
   x2E80AC30 : -> Block
   x2FD76FFB : -> Block
   x30261492 : -> Block
   x303FF4AA : -> Block
   x33D5A466 : -> Block
   x344925FC : -> Block
   x34ACF886 : -> Block
   x3CD54DEB : -> Block
   x3CF3A7D2 : -> Block
   x3DD81AC6 : -> Block
   x3F6F7248 : -> Block
   x48B204D6 : -> Block
   x4A645A01 : -> Block
   x4C49AAE0 : -> Block
   x4CE933E1 : -> Block
   x4D53901A : -> Block
   x4DA124A1 : -> Block
   x4F998E01 : -> Block
   x4FB1138A : -> Block
   x50DEC930 : -> Block
   x51AF3C1D : -> Block
   x51EDE9C7 : -> Block
   x550D91CE : -> Block
   x55555555 : -> Block
   x55DD063F : -> Block
   x5834A585 : -> Block
   x5A35D667 : -> Block
   x5BC02502 : -> Block
   x5CCA3239 : -> Block
   x5EBA06C2 : -> Block
   x5F38EEF1 : -> Block
   x613F8E2A : -> Block
   x63C70DBA : -> Block
   x6AD6E8A4 : -> Block
   x6AEBACF8 : -> Block
   x6D67E884 : -> Block
   x7050EC5E : -> Block
   x717153D5 : -> Block
   x7201F4DC : -> Block
   x7397C9AE : -> Block
   x74B39176 : -> Block
   x76232E5F : -> Block
   x7783C51D : -> Block
   x7792F9D4 : -> Block
   x7BC180AB : -> Block
   x7DB2D9F4 : -> Block
   x7DFEFBFF : -> Block   % MAA special constant 'D'
   x7F76A3B0 : -> Block
   x7F839576 : -> Block
   x7FFFFFF0 : -> Block
   x7FFFFFF1 : -> Block
   x7FFFFFFC : -> Block
   x7FFFFFFD : -> Block
   x80000000 : -> Block
   x80000002 : -> Block
   x800000C2 : -> Block
   x80018000 : -> Block
   x80018001 : -> Block
   x80397302 : -> Block
   x81D10CA3 : -> Block
   x89D635D7 : -> Block
   x8CE37709 : -> Block
   x8DC8BBDE : -> Block
   x9115A558 : -> Block
   x91896CFA : -> Block
   x9372CDC6 : -> Block
   x98D1CC75 : -> Block
   x9D15C437 : -> Block
   x9DB15CF6 : -> Block
   x9E2E7B36 : -> Block
   xA018C83B : -> Block
   xA0B87B77 : -> Block
   xA44AAAC0 : -> Block
   xA511987A : -> Block
   xA70FC148 : -> Block
   xA93BD410 : -> Block
   xAAAAAAAA : -> Block
   xAB00FFCD : -> Block
   xAB01FCCD : -> Block
   xAB6EED4A : -> Block
   xABEEED6B : -> Block
   xACBC13DD : -> Block
   xB1CC1CC5 : -> Block
   xB8142629 : -> Block
   xB99A62DE : -> Block
   xBA92DB12 : -> Block
   xBBA57835 : -> Block
   xBE9F0917 : -> Block
   xBF2D7D85 : -> Block
   xBFEF7FDF : -> Block   % MAA special constant 'C'
   xC1ED90DD : -> Block
   xC21A1846 : -> Block
   xC4EB1AEB : -> Block
   xC6B1317E : -> Block
   xCBC865BA : -> Block
   xCD959B46 : -> Block
   xD0482465 : -> Block
   xD636250D : -> Block
   xD7843FDC : -> Block
   xD78634BC : -> Block
   xD8804CA5 : -> Block
   xDB79FBDC : -> Block
   xDB9102B0 : -> Block
   xE0C08000 : -> Block
   xE6A12F07 : -> Block
   xEB35B97F : -> Block
   xF0239DD5 : -> Block
   xF14D6E28 : -> Block
   xF2EF3501 : -> Block
   xF6A09667 : -> Block
   xFD297DA4 : -> Block
   xFDC1A8BA : -> Block
   xFE4E5BDD : -> Block
   xFEA1D334 : -> Block
   xFECCAA6E : -> Block
   xFEFC07F0 : -> Block
   xFF2D7DA5 : -> Block
   xFFEF0001 : -> Block
   xFFFF00FF : -> Block
   xFFFFFF2D : -> Block
   xFFFFFF3A : -> Block
   xFFFFFFF0 : -> Block
   xFFFFFFF1 : -> Block
   xFFFFFFF4 : -> Block
   xFFFFFFF5 : -> Block
   xFFFFFFF7 : -> Block
   xFFFFFFF9 : -> Block
   xFFFFFFFA : -> Block
   xFFFFFFFB : -> Block
   xFFFFFFFC : -> Block
   xFFFFFFFD : -> Block
   xFFFFFFFE : -> Block
   xFFFFFFFF : -> Block
 VARS
   O1 O2 O3 O4 O'1 O'2 O'3 O'4 O"1 O"2 O"3 O"4 : Octet
   O11U O11L O12U O12L O21U O21L O22U O22L Ocarry : Octet
 RULES
   eqBlock (buildBlock (O1, O2, O3, O4), buildBlock (O'1, O'2, O'3, O'4))
   -> andBool (andBool (eqOctet (O1, O'1), eqOctet (O2, O'2)),
               andBool (eqOctet (O3, O'3), eqOctet (O4, O'4)))

   andBlock (buildBlock (O1, O2, O3, O4), buildBlock (O'1, O'2, O'3, O'4))
   -> buildBlock (andOctet (O1, O'1), andOctet (O2, O'2),
                  andOctet (O3, O'3), andOctet (O4, O'4))

   orBlock (buildBlock (O1, O2, O3, O4), buildBlock (O'1, O'2, O'3, O'4))
   -> buildBlock (orOctet (O1, O'1), orOctet (O2, O'2),
                  orOctet (O3, O'3), orOctet (O4, O'4))

   xorBlock (buildBlock (O1, O2, O3, O4), buildBlock (O'1, O'2, O'3, O'4))
   -> buildBlock (xorOctet (O1, O'1), xorOctet (O2, O'2),
                  xorOctet (O3, O'3), xorOctet (O4, O'4))

   HalfU (buildBlock (O1, O2, O3, O4)) -> buildHalf (O1, O2)

   HalfL (buildBlock (O1, O2, O3, O4)) -> buildHalf (O3, O4)

   mulHalf (buildHalf (O1, O2), buildHalf (O'1, O'2))
   -> mulHalfA (mulOctet (O1, O'1), mulOctet (O1, O'2),
                mulOctet (O2, O'1), mulOctet (O2, O'2))

   mulHalfA (buildHalf (O11U, O11L), buildHalf (O12U, O12L),
             buildHalf (O21U, O21L), buildHalf (O22U, O22L))
   -> mulHalf4 (O11U, O11L, O12U, O12L, O21U, O21L; O22U; O22L)

   mulHalf4 (O11U, O11L, O12U, O12L, O21U, O21L; O22U; O"4)
   -> mulHalf3 (O11U, O11L, O12U, O21U;
                addHalfOctet (O12L, addHalfOctets (O21L, O22U)); O"4)

   mulHalf3 (O11U, O11L, O12U, O21U; buildHalf (Ocarry, O"3); O"4)
   -> mulHalf2 (O11U; addHalfOctet (Ocarry,
                      addHalfOctet (O11L, addHalfOctets (O12U, O21U))); O"3, O"4)

   mulHalf2 (O11U; buildHalf (Ocarry, O"2); O"3, O"4)
   -> mulHalf1 (addHalfOctets (Ocarry, O11U); O"2; O"3, O"4)

   mulHalf1 (buildHalf (Ocarry, O"1); O"2; O"3, O"4)
   -> buildBlock (O"1, O"2, O"3, O"4) % assert eqOctet (Ocarry, x00)

   x00000000 -> buildBlock (x00, x00, x00, x00)
   x00000001 -> buildBlock (x00, x00, x00, x01)
   x00000002 -> buildBlock (x00, x00, x00, x02)
   x00000003 -> buildBlock (x00, x00, x00, x03)
   x00000004 -> buildBlock (x00, x00, x00, x04)
   x00000005 -> buildBlock (x00, x00, x00, x05)
   x00000006 -> buildBlock (x00, x00, x00, x06)
   x00000007 -> buildBlock (x00, x00, x00, x07)
   x00000008 -> buildBlock (x00, x00, x00, x08)
   x00000009 -> buildBlock (x00, x00, x00, x09)
   x0000000A -> buildBlock (x00, x00, x00, x0A)
   x0000000B -> buildBlock (x00, x00, x00, x0B)
   x0000000C -> buildBlock (x00, x00, x00, x0C)
   x0000000D -> buildBlock (x00, x00, x00, x0D)
   x0000000E -> buildBlock (x00, x00, x00, x0E)
   x0000000F -> buildBlock (x00, x00, x00, x0F)
   x00000010 -> buildBlock (x00, x00, x00, x10)
   x00000012 -> buildBlock (x00, x00, x00, x12)
   x00000014 -> buildBlock (x00, x00, x00, x14)
   x00000016 -> buildBlock (x00, x00, x00, x16)
   x00000018 -> buildBlock (x00, x00, x00, x18)
   x0000001B -> buildBlock (x00, x00, x00, x1B)
   x0000001D -> buildBlock (x00, x00, x00, x1D)
   x0000001E -> buildBlock (x00, x00, x00, x1E)
   x0000001F -> buildBlock (x00, x00, x00, x1F)
   x00000031 -> buildBlock (x00, x00, x00, x31)
   x00000036 -> buildBlock (x00, x00, x00, x36)
   x00000060 -> buildBlock (x00, x00, x00, x60)
   x00000080 -> buildBlock (x00, x00, x00, x80)
   x000000A5 -> buildBlock (x00, x00, x00, xA5)
   x000000B6 -> buildBlock (x00, x00, x00, xB6)
   x000000C4 -> buildBlock (x00, x00, x00, xC4)
   x000000D2 -> buildBlock (x00, x00, x00, xD2)
   x00000100 -> buildBlock (x00, x00, x01, x00)
   x00000129 -> buildBlock (x00, x00, x01, x29)
   x0000018C -> buildBlock (x00, x00, x01, x8C)
   x00004000 -> buildBlock (x00, x00, x40, x00)
   x00010000 -> buildBlock (x00, x01, x00, x00)
   x00020000 -> buildBlock (x00, x02, x00, x00)
   x00030000 -> buildBlock (x00, x03, x00, x00)
   x00040000 -> buildBlock (x00, x04, x00, x00)
   x00060000 -> buildBlock (x00, x06, x00, x00)
   x00804021 -> buildBlock (x00, x80, x40, x21)   % MAA special constant 'B'
   x00FF00FF -> buildBlock (x00, xFF, x00, xFF)
   x0103050B -> buildBlock (x01, x03, x05, x0B)
   x01030703 -> buildBlock (x01, x03, x07, x03)
   x01030705 -> buildBlock (x01, x03, x07, x05)
   x0103070F -> buildBlock (x01, x03, x07, x0F)
   x02040801 -> buildBlock (x02, x04, x08, x01)   % MAA special constant 'A'
   x0297AF6F -> buildBlock (x02, x97, xAF, x6F)
   x07050301 -> buildBlock (x07, x05, x03, x01)
   x077788A2 -> buildBlock (x07, x77, x88, xA2)
   x07C72EAA -> buildBlock (x07, xC7, x2E, xAA)
   x0A202020 -> buildBlock (x0A, x20, x20, x20)
   x0AD67E20 -> buildBlock (x0A, xD6, x7E, x20)
   x10000000 -> buildBlock (x10, x00, x00, x00)
   x11A9D254 -> buildBlock (x11, xA9, xD2, x54)
   x11AC46B8 -> buildBlock (x11, xAC, x46, xB8)
   x1277A6D4 -> buildBlock (x12, x77, xA6, xD4)
   x13647149 -> buildBlock (x13, x64, x71, x49)
   x160EE9B5 -> buildBlock (x16, x0E, xE9, xB5)
   x17065DBB -> buildBlock (x17, x06, x5D, xBB)
   x17A808FD -> buildBlock (x17, xA8, x08, xFD)
   x1D10D8D3 -> buildBlock (x1D, x10, xD8, xD3)
   x1D3B7760 -> buildBlock (x1D, x3B, x77, x60)
   x1D9C9655 -> buildBlock (x1D, x9C, x96, x55)
   x1F3F7FFF -> buildBlock (x1F, x3F, x7F, xFF)
   x204E80A7 -> buildBlock (x20, x4E, x80, xA7)
   x21D869BA -> buildBlock (x21, xD8, x69, xBA)
   x24B66FB5 -> buildBlock (x24, xB6, x6F, xB5)
   x270EEDAF -> buildBlock (x27, x0E, xED, xAF)
   x277B4B25 -> buildBlock (x27, x7B, x4B, x25)
   x2829040B -> buildBlock (x28, x29, x04, x0B)
   x288FC786 -> buildBlock (x28, x8F, xC7, x86)
   x28EAD8B3 -> buildBlock (x28, xEA, xD8, xB3)
   x29907CD8 -> buildBlock (x29, x90, x7C, xD8)
   x29C1485F -> buildBlock (x29, xC1, x48, x5F)
   x29EEE96B -> buildBlock (x29, xEE, xE9, x6B)
   x2A6091AE -> buildBlock (x2A, x60, x91, xAE)
   x2BF8499A -> buildBlock (x2B, xF8, x49, x9A)
   x2E80AC30 -> buildBlock (x2E, x80, xAC, x30)
   x2FD76FFB -> buildBlock (x2F, xD7, x6F, xFB)
   x30261492 -> buildBlock (x30, x26, x14, x92)
   x303FF4AA -> buildBlock (x30, x3F, xF4, xAA)
   x33D5A466 -> buildBlock (x33, xD5, xA4, x66)
   x344925FC -> buildBlock (x34, x49, x25, xFC)
   x34ACF886 -> buildBlock (x34, xAC, xF8, x86)
   x3CD54DEB -> buildBlock (x3C, xD5, x4D, xEB)
   x3CF3A7D2 -> buildBlock (x3C, xF3, xA7, xD2)
   x3DD81AC6 -> buildBlock (x3D, xD8, x1A, xC6)
   x3F6F7248 -> buildBlock (x3F, x6F, x72, x48)
   x48B204D6 -> buildBlock (x48, xB2, x04, xD6)
   x4A645A01 -> buildBlock (x4A, x64, x5A, x01)
   x4C49AAE0 -> buildBlock (x4C, x49, xAA, xE0)
   x4CE933E1 -> buildBlock (x4C, xE9, x33, xE1)
   x4D53901A -> buildBlock (x4D, x53, x90, x1A)
   x4DA124A1 -> buildBlock (x4D, xA1, x24, xA1)
   x4F998E01 -> buildBlock (x4F, x99, x8E, x01)
   x4FB1138A -> buildBlock (x4F, xB1, x13, x8A)
   x50DEC930 -> buildBlock (x50, xDE, xC9, x30)
   x51AF3C1D -> buildBlock (x51, xAF, x3C, x1D)
   x51EDE9C7 -> buildBlock (x51, xED, xE9, xC7)
   x550D91CE -> buildBlock (x55, x0D, x91, xCE)
   x55555555 -> buildBlock (x55, x55, x55, x55)
   x55DD063F -> buildBlock (x55, xDD, x06, x3F)
   x5834A585 -> buildBlock (x58, x34, xA5, x85)
   x5A35D667 -> buildBlock (x5A, x35, xD6, x67)
   x5BC02502 -> buildBlock (x5B, xC0, x25, x02)
   x5CCA3239 -> buildBlock (x5C, xCA, x32, x39)
   x5EBA06C2 -> buildBlock (x5E, xBA, x06, xC2)
   x5F38EEF1 -> buildBlock (x5F, x38, xEE, xF1)
   x613F8E2A -> buildBlock (x61, x3F, x8E, x2A)
   x63C70DBA -> buildBlock (x63, xC7, x0D, xBA)
   x6AD6E8A4 -> buildBlock (x6A, xD6, xE8, xA4)
   x6AEBACF8 -> buildBlock (x6A, xEB, xAC, xF8)
   x6D67E884 -> buildBlock (x6D, x67, xE8, x84)
   x7050EC5E -> buildBlock (x70, x50, xEC, x5E)
   x717153D5 -> buildBlock (x71, x71, x53, xD5)
   x7201F4DC -> buildBlock (x72, x01, xF4, xDC)
   x7397C9AE -> buildBlock (x73, x97, xC9, xAE)
   x74B39176 -> buildBlock (x74, xB3, x91, x76)
   x76232E5F -> buildBlock (x76, x23, x2E, x5F)
   x7783C51D -> buildBlock (x77, x83, xC5, x1D)
   x7792F9D4 -> buildBlock (x77, x92, xF9, xD4)
   x7BC180AB -> buildBlock (x7B, xC1, x80, xAB)
   x7DB2D9F4 -> buildBlock (x7D, xB2, xD9, xF4)
   x7DFEFBFF -> buildBlock (x7D, xFE, xFB, xFF)   % MAA special constant 'D'
   x7F76A3B0 -> buildBlock (x7F, x76, xA3, xB0)
   x7F839576 -> buildBlock (x7F, x83, x95, x76)
   x7FFFFFF0 -> buildBlock (x7F, xFF, xFF, xF0)
   x7FFFFFF1 -> buildBlock (x7F, xFF, xFF, xF1)
   x7FFFFFFC -> buildBlock (x7F, xFF, xFF, xFC)
   x7FFFFFFD -> buildBlock (x7F, xFF, xFF, xFD)
   x80000000 -> buildBlock (x80, x00, x00, x00)
   x80000002 -> buildBlock (x80, x00, x00, x02)
   x800000C2 -> buildBlock (x80, x00, x00, xC2)
   x80018000 -> buildBlock (x80, x01, x80, x00)
   x80018001 -> buildBlock (x80, x01, x80, x01)
   x80397302 -> buildBlock (x80, x39, x73, x02)
   x81D10CA3 -> buildBlock (x81, xD1, x0C, xA3)
   x89D635D7 -> buildBlock (x89, xD6, x35, xD7)
   x8CE37709 -> buildBlock (x8C, xE3, x77, x09)
   x8DC8BBDE -> buildBlock (x8D, xC8, xBB, xDE)
   x9115A558 -> buildBlock (x91, x15, xA5, x58)
   x91896CFA -> buildBlock (x91, x89, x6C, xFA)
   x9372CDC6 -> buildBlock (x93, x72, xCD, xC6)
   x98D1CC75 -> buildBlock (x98, xD1, xCC, x75)
   x9D15C437 -> buildBlock (x9D, x15, xC4, x37)
   x9DB15CF6 -> buildBlock (x9D, xB1, x5C, xF6)
   x9E2E7B36 -> buildBlock (x9E, x2E, x7B, x36)
   xA018C83B -> buildBlock (xA0, x18, xC8, x3B)
   xA0B87B77 -> buildBlock (xA0, xB8, x7B, x77)
   xA44AAAC0 -> buildBlock (xA4, x4A, xAA, xC0)
   xA511987A -> buildBlock (xA5, x11, x98, x7A)
   xA70FC148 -> buildBlock (xA7, x0F, xC1, x48)
   xA93BD410 -> buildBlock (xA9, x3B, xD4, x10)
   xAAAAAAAA -> buildBlock (xAA, xAA, xAA, xAA)
   xAB00FFCD -> buildBlock (xAB, x00, xFF, xCD)
   xAB01FCCD -> buildBlock (xAB, x01, xFC, xCD)
   xAB6EED4A -> buildBlock (xAB, x6E, xED, x4A)
   xABEEED6B -> buildBlock (xAB, xEE, xED, x6B)
   xACBC13DD -> buildBlock (xAC, xBC, x13, xDD)
   xB1CC1CC5 -> buildBlock (xB1, xCC, x1C, xC5)
   xB8142629 -> buildBlock (xB8, x14, x26, x29)
   xB99A62DE -> buildBlock (xB9, x9A, x62, xDE)
   xBA92DB12 -> buildBlock (xBA, x92, xDB, x12)
   xBBA57835 -> buildBlock (xBB, xA5, x78, x35)
   xBE9F0917 -> buildBlock (xBE, x9F, x09, x17)
   xBF2D7D85 -> buildBlock (xBF, x2D, x7D, x85)
   xBFEF7FDF -> buildBlock (xBF, xEF, x7F, xDF)   % MAA special constant 'C'
   xC1ED90DD -> buildBlock (xC1, xED, x90, xDD)
   xC21A1846 -> buildBlock (xC2, x1A, x18, x46)
   xC4EB1AEB -> buildBlock (xC4, xEB, x1A, xEB)
   xC6B1317E -> buildBlock (xC6, xB1, x31, x7E)
   xCBC865BA -> buildBlock (xCB, xC8, x65, xBA)
   xCD959B46 -> buildBlock (xCD, x95, x9B, x46)
   xD0482465 -> buildBlock (xD0, x48, x24, x65)
   xD636250D -> buildBlock (xD6, x36, x25, x0D)
   xD7843FDC -> buildBlock (xD7, x84, x3F, xDC)
   xD78634BC -> buildBlock (xD7, x86, x34, xBC)
   xD8804CA5 -> buildBlock (xD8, x80, x4C, xA5)
   xDB79FBDC -> buildBlock (xDB, x79, xFB, xDC)
   xDB9102B0 -> buildBlock (xDB, x91, x02, xB0)
   xE0C08000 -> buildBlock (xE0, xC0, x80, x00)
   xE6A12F07 -> buildBlock (xE6, xA1, x2F, x07)
   xEB35B97F -> buildBlock (xEB, x35, xB9, x7F)
   xF0239DD5 -> buildBlock (xF0, x23, x9D, xD5)
   xF14D6E28 -> buildBlock (xF1, x4D, x6E, x28)
   xF2EF3501 -> buildBlock (xF2, xEF, x35, x01)
   xF6A09667 -> buildBlock (xF6, xA0, x96, x67)
   xFD297DA4 -> buildBlock (xFD, x29, x7D, xA4)
   xFDC1A8BA -> buildBlock (xFD, xC1, xA8, xBA)
   xFE4E5BDD -> buildBlock (xFE, x4E, x5B, xDD)
   xFEA1D334 -> buildBlock (xFE, xA1, xD3, x34)
   xFECCAA6E -> buildBlock (xFE, xCC, xAA, x6E)
   xFEFC07F0 -> buildBlock (xFE, xFC, x07, xF0)
   xFF2D7DA5 -> buildBlock (xFF, x2D, x7D, xA5)
   xFFEF0001 -> buildBlock (xFF, xEF, x00, x01)
   xFFFF00FF -> buildBlock (xFF, xFF, x00, xFF)
   xFFFFFF2D -> buildBlock (xFF, xFF, xFF, x2D)
   xFFFFFF3A -> buildBlock (xFF, xFF, xFF, x3A)
   xFFFFFFF0 -> buildBlock (xFF, xFF, xFF, xF0)
   xFFFFFFF1 -> buildBlock (xFF, xFF, xFF, xF1)
   xFFFFFFF4 -> buildBlock (xFF, xFF, xFF, xF4)
   xFFFFFFF5 -> buildBlock (xFF, xFF, xFF, xF5)
   xFFFFFFF7 -> buildBlock (xFF, xFF, xFF, xF7)
   xFFFFFFF9 -> buildBlock (xFF, xFF, xFF, xF9)
   xFFFFFFFA -> buildBlock (xFF, xFF, xFF, xFA)
   xFFFFFFFB -> buildBlock (xFF, xFF, xFF, xFB)
   xFFFFFFFC -> buildBlock (xFF, xFF, xFF, xFC)
   xFFFFFFFD -> buildBlock (xFF, xFF, xFF, xFD)
   xFFFFFFFE -> buildBlock (xFF, xFF, xFF, xFE)
   xFFFFFFFF -> buildBlock (xFF, xFF, xFF, xFF)
\end{verbatim}
\end{small}

\subsection{Definitions for sort BlockSum}

We now define sort \T{BlockSum} that stores the result of the addition of two blocks. Values of this sort are 33-bit words, made up using the constructor \T{buildBlockSum} that gathers one bit for the carry and a block for the sum. The five principal non-constructors for this sort are \T{eqBlockSum} (which tests equality), \T{addBlockSum} (which adds two blocks and returns both a carry bit and a 32-bit sum), \T{addBlock} (which is derived from the former one by dropping the carry bit), \T{addBlockHalf} and \T{addBlockHalves} (which are similar to the former one but take half-word arguments that are converted to blocks before summation); the other non-constructors are auxiliary functions implementing a 32-bit adder built using four 8-bit adders.

\begin{small}
\begin{verbatim}
 SORTS
   BlockSum
 CONS
   buildBlockSum : Bit Block -> BlockSum
 OPNS
   eqBlockSum : BlockSum BlockSum -> Bool
   addBlockSum : Block Block -> BlockSum
   addBlock4 : Octet Octet Octet Octet Octet Octet Octet Octet -> BlockSum
   addBlock3 : Octet Octet Octet Octet Octet Octet OctetSum -> BlockSum
   addBlock2 : Octet Octet Octet Octet OctetSum Octet -> BlockSum
   addBlock1 : Octet Octet OctetSum Octet Octet -> BlockSum
   addBlock0 : OctetSum Octet Octet Octet -> BlockSum
   dropCarryBlockSum : BlockSum -> Block
   addBlock : Block Block -> Block
   addBlockHalf : Half Block -> Block
   addBlockHalves : Half Half -> Block
 VARS
   B B' Bcarry : Bit
   O1 O2 O3 O4 O'1 O'2 O'3 O'4 O"1 O"2 O"3 O"4 : Octet
   W W' : Block
 RULES
   eqBlockSum (buildBlockSum (B, W), buildBlockSum (B', W'))
   -> andBool (eqBit (B, B'), eqBlock (W, W'))

   addBlockSum (buildBlock (O1, O2, O3, O4), buildBlock (O'1, O'2, O'3, O'4))
   -> addBlock4 (O1, O'1, O2, O'2, O3, O'3, O4, O'4)

   addBlock4 (O1, O'1, O2, O'2, O3, O'3, O4, O'4)
   -> addBlock3 (O1, O'1, O2, O'2, O3, O'3, addOctetSum (O4, O'4, x0))

   addBlock3 (O1, O'1, O2, O'2, O3, O'3, buildOctetSum (Bcarry, O"4))
   -> addBlock2 (O1, O'1, O2, O'2, addOctetSum (O3, O'3, Bcarry); O"4)

   addBlock2 (O1, O'1, O2, O'2, buildOctetSum (Bcarry, O"3); O"4)
   -> addBlock1 (O1, O'1, addOctetSum (O2, O'2, Bcarry); O"3, O"4)

   addBlock1 (O1, O'1, buildOctetSum (Bcarry, O"2); O"3, O"4)
   -> addBlock0 (addOctetSum (O1, O'1, Bcarry); O"2, O"3, O"4)

   addBlock0 (buildOctetSum (Bcarry, O"1); O"2, O"3, O"4)
   -> buildBlockSum (Bcarry, buildBlock (O"1, O"2, O"3, O"4))

   dropCarryBlockSum (buildBlockSum (Bcarry, W)) -> W

   addBlock (W, W') -> dropCarryBlockSum (addBlockSum (W, W'))

   addBlockHalf (buildHalf (O1, O2), W)
   -> addBlock (buildBlock (x00, x00, O1, O2), W)

   addBlockHalves (buildHalf (O1, O2), buildHalf (O'1, O'2))
   -> addBlock (buildBlock (x00, x00, O1, O2), buildBlock (x00, x00, O'1, O'2))
\end{verbatim}
\end{small}

\subsection{Definitions for sort Pair}

We now define 64-bit words (named ``pairs'' according to the MAA terminology) using a constructor \T{buildPair} that takes two blocks and returns a pair. The two principal non-constructors for this sort are \T{eqPair} (which tests equality) and \T{mulBlock} (which takes two blocks and computes their 64-bit product); the other non-constructors are auxiliary functions implementing a 32-bit multiplier built using four 16-bit multipliers.

\begin{small}
\begin{verbatim}
 SORTS
   Pair
 CONS
   buildPair : Block Block -> Pair
   % the first argument of buildPair contain the most significant bits
 OPNS
   eqPair : Pair Pair -> Bool
   mulBlock : Block Block -> Pair
   mulBlockA : Block Block Block Block -> Pair
   mulBlock4 : Half Half Half Half Half Half Half Half -> Pair
   mulBlock3 : Half Half Half Half Block Half -> Pair
   mulBlock2 : Half Block Half Half -> Pair
   mulBlock1 : Block Half Half Half -> Pair
   mulBlockB : Half Half Half Half -> Pair
 VARS
   O1 O2 O3 O4 O'1 O'2 O'3 O'4 : Octet
   O1U O1L O2U O2L O3U O3L O4U O4L : Octet
   H"2 H"3 H"4 : Half
   H11U H11L H12U H12L H21U H21L H22U H22L : Half
   W W1 W2 W'1 W'2 : Block
   W11 W12 W21 W22 : Block
 RULES
   eqPair (buildPair (W1, W2), buildPair (W'1, W'2))
   -> andBool (eqBlock (W1, W'1), eqBlock (W2, W'2))

   mulBlock (W1, W2)
   -> mulBlockA (mulHalf (HalfU (W1), HalfU (W2)), mulHalf (HalfU (W1), HalfL (W2)),
                 mulHalf (HalfL (W1), HalfU (W2)), mulHalf (HalfL (W1), HalfL (W2)))

   mulBlockA (W11, W12, W21, W22)
   -> mulBlock4 (HalfU (W11), HalfL (W11), HalfU (W12), HalfL (W12),
                 HalfU (W21), HalfL (W21); HalfU (W22); HalfL (W22))

  mulBlock4 (H11U, H11L, H12U, H12L, H21U, H21L; H22U; H"4)
  -> mulBlock3 (H11U, H11L, H12U, H21U;
                addBlockHalf (H12L, addBlockHalves (H21L, H22U)); H"4)

  mulBlock3 (H11U, H11L, H12U, H21U; W; H"4)
  -> mulBlock2 (H11U; addBlockHalf (HalfU (W),
                addBlockHalf (H11L, addBlockHalves (H12U, H21U))); HalfL (W), H"4)

  mulBlock2 (H11U; W; H"3, H"4)
  -> mulBlock1 (addBlockHalves (HalfU (W), H11U); HalfL (W), H"3, H"4)

  mulBlock1 (W; H"2, H"3, H"4)
  -> mulBlockB (HalfL (W), H"2, H"3, H"4) % assert eqHalf (HalfU (W), x0000)

  mulBlockB (buildHalf (O1U, O1L), buildHalf (O2U, O2L),
             buildHalf (O3U, O3L), buildHalf (O4U, O4L))
  -> buildPair (buildBlock (O1U, O1L, O2U, O2L), buildBlock (O3U, O3L, O4U, O4L))
\end{verbatim}
\end{small}

\subsection{Definitions for sort Key}
\label{ANNEX-REC-SPECIFIC}

We now define a sort \T{Key} that is intended to represent the 64-bit keys $(J, K)$ used by the MAA. This sort has a constructor \T{buildKey} that takes two blocks and returns a key. In \cite{Munster-91-a}, keys are represented using the sort \T{Pair}, but we prefer introducing a dedicated sort to clearly distinguish between keys and, e.g., results of the multiplication of two blocks.

\begin{small}
\begin{verbatim}
 SORTS
   Key
 CONS
   buildKey : Block Block -> Key
   % the 1st argument of buildKey was noted J in the MAA specification
   % the 2nd argument of buildKey was noted K in the MAA specification
\end{verbatim}
\end{small}

\subsection{Definitions for sort Message}
\label{ANNEX-REC-MESSAGE}

We now define messages, which are non-empty lists of blocks built using two constructors \T{unitMessage} and \T{consMessage}; there are three non-constructors for this sort: \T{appendMessage} (which inserts a block at the end of a list), \T{reverseMessage} (which reverses a list), and \T{makeMessage} (which generates a message of a given length, the blocks of which follow an arithmetic progression).

\begin{small}
\begin{verbatim}
 SORTS
   Message
 CONS
   unitMessage : Block -> Message
   consMessage : Block Message -> Message
 OPNS
   appendMessage : Message Block -> Message
   reverseMessage : Message -> Message
   makeMessage : Nat Block Block -> Message
 VARS
   M M' : Message
   W W' : Block
 RULES
   appendMessage (unitMessage (W), W') -> consMessage (W, unitMessage (W'))
   appendMessage (consMessage (W, M), W') -> consMessage (W, appendMessage (M, W'))

   reverseMessage (unitMessage (W)) -> unitMessage (W)
   reverseMessage (consMessage (W, M)) -> appendMessage (reverseMessage (M), W)

   makeMessage (succ (N), W, W')
   -> unitMessage (W) if eqNat (N, zero) -><- true
   makeMessage (succ (N), W, W')
   -> consMessage (W, makeMessage (N, ADD (W, W'), W')) if eqNat (N, zero) -><- false
\end{verbatim}
\end{small}

\noindent If needed, the two conditional rules could be eliminated by modifying the definition of \T{makeMessage} as follows:

\begin{small}
\begin{verbatim}
   makeMessage (succ (zero), W, W')
   -> unitMessage (W)
   makeMessage (succ (succ (N)), W, W')
   -> consMessage (W, makeMessage (succ (N), ADD (W, W'), W'))
\end{verbatim}
\end{small}

\subsection{Definitions for sort SegmentedMessage}
\label{SEGMENTS}

We now define segmented messages, which are non-empty lists of messages, each message containing up to 1204 octets (i.e., 256 blocks). Values of this sort are built using two constructors \T{unitSegment} and \T{consSegment}; the principal non-constructor is \T{splitSegment}, which converts a message into a segmented message.

\begin{small}
\begin{verbatim}
 SORTS
   SegmentedMessage
 CONS
   unitSegment : Message -> SegmentedMessage
   consSegment : Message SegmentedMessage -> SegmentedMessage
 OPNS
   splitSegment : Message -> SegmentedMessage
   cutSegment : Message Message Nat -> SegmentedMessage
 VARS
   M M' : Message
   N : Nat
   S : SegmentedMessage
   W : Block
 RULES
   splitSegment (unitMessage (W)) -> unitSegment (unitMessage (W))
   splitSegment (consMessage (W, M)) -> cutSegment (M, unitMessage (W), n254)

   cutSegment (unitMessage (W), M', N)
   -> unitSegment (reverseMessage (consMessage (W, M')))
   cutSegment (consMessage (W, M), M', zero)
   -> consSegment (reverseMessage (consMessage (W, M')), splitSegment (M))
   cutSegment (consMessage (W, M), M', succ (N))
   -> cutSegment (M, consMessage (W, M'), N)
\end{verbatim}
\end{small}

\subsection{Definitions (1) of MAA-specific cryptographic functions}

We now define a first set of functions to be used for MAA computations, most of which were present in \cite{Davies-Clayden-88} or have been later introduced in \cite{Munster-91-a}. Operations \T{ADD}, \T{AND}, \T{MUL}, \T{OR}, and \T{XOR} are merely aliases of already-defined functions on Blocks; operations \T{BYT'} and \T{ADDC'} are just auxiliary functions.

\begin{small}
\begin{verbatim}
 OPNS
   ADD : Block Block -> Block
   AND : Block Block -> Block
   MUL : Block Block -> Pair
   OR : Block Block -> Block
   XOR : Block Block -> Block
   XOR' : Pair -> Block
   CYC : Block -> Block
   nCYC : Nat Block -> Block
   FIX1 : Block -> Block
   FIX2 : Block -> Block
   needAdjust : Octet -> Bool
   adjustCode : Octet -> Bit
   adjust : Octet Octet -> Octet
   PAT : Block Block -> Octet
   BYT : Block Block -> Pair
   BYT' : Octet Octet Octet Octet Octet Octet Octet Octet Octet -> Pair
   ADDC : Block Block -> Pair
   ADDC' : BlockSum -> Pair
 VARS
   B1 B2 B3 B4 B5 B6 B7 B8 : Bit
   B9 B10 B11 B12 B13 B14 B15 B16 : Bit
   B17 B18 B19 B20 B21 B22 B23 B24 : Bit
   B25 B26 B27 B28 B29 B30 B31 B32 : Bit
   O O' : Octet
   W W' : Block
 RULES
   ADD (W, W') -> addBlock (W, W')

   AND (W, W') -> andBlock (W, W')

   MUL (W, W') -> mulBlock (W, W')

   OR (W, W') -> orBlock (W, W')

   XOR (W, W') -> xorBlock (W, W')

   XOR' (buildPair (W, W')) -> XOR (W, W')

   CYC (buildBlock (buildOctet (B1, B2, B3, B4, B5, B6, B7, B8),
                    buildOctet (B9, B10, B11, B12, B13, B14, B15, B16),
                    buildOctet (B17, B18, B19, B20, B21, B22, B23, B24),
                    buildOctet (B25, B26, B27, B28, B29, B30, B31, B32)))
   -> buildBlock (buildOctet (B2, B3, B4, B5, B6, B7, B8, B9),
                  buildOctet (B10, B11, B12, B13, B14, B15, B16, B17),
                  buildOctet (B18, B19, B20, B21, B22, B23, B24, B25),
                  buildOctet (B26, B27, B28, B29, B30, B31, B32, B1))

   nCYC (zero, W) -> W
   nCYC (succ (N), W) -> CYC (nCYC (N, W))

   FIX1 (W) -> AND (OR (W, x02040801), xBFEF7FDF)   % A = x02040801, C = xBFEF7FDF

   FIX2 (W) -> AND (OR (W, x00804021), x7DFEFBFF)   % B = x00804021, D = x7DFEFBFF

   needAdjust (O) -> orBool (eqOctet (O, x00), eqOctet (O, xFF))

   adjustCode (O) -> x1               if needAdjust (O) -><- true
   adjustCode (O) -> x0               if needAdjust (O) -><- false

   adjust (O, O') -> xorOctet (O, O') if needAdjust (O) -><- true
   adjust (O, O') -> O                if needAdjust (O) -><- false

   PAT (buildBlock (O1, O2, O3, O4), buildBlock (O'1, O'2, O'3, O'4))
   -> buildOctet (adjustCode (O1), adjustCode (O2),
                  adjustCode (O3), adjustCode (O4),
                  adjustCode (O'1), adjustCode (O'2),
                  adjustCode (O'3), adjustCode (O'4))

   BYT (buildBlock (O1, O2, O3, O4), buildBlock (O'1, O'2, O'3, O'4))
   -> BYT' (O1, O2, O3, O4, O'1, O'2, O'3, O'4,
            PAT (buildBlock (O1, O2, O3, O4), buildBlock (O'1, O'2, O'3, O'4)))

   BYT' (O1, O2, O3, O4, O'1, O'2, O'3, O'4, Opat)
   -> buildPair (buildBlock (adjust (O1, rightOctet7 (Opat)),
                             adjust (O2, rightOctet6 (Opat)),
                             adjust (O3, rightOctet5 (Opat)),
                             adjust (O4, rightOctet4 (Opat))),
                 buildBlock (adjust (O'1, rightOctet3 (Opat)),
                             adjust (O'2, rightOctet2 (Opat)),
                             adjust (O'3, rightOctet1 (Opat)),
                             adjust (O'4, Opat)))

   ADDC (W, W') -> ADDC' (addBlockSum (W, W'))

   ADDC' (buildBlockSum (x0, W)) -> buildPair (x00000000, W)
   ADDC' (buildBlockSum (x1, W)) -> buildPair (x00000001, W)
\end{verbatim}
\end{small}

\noindent If needed, the four conditional rules could be eliminated by introducing two auxiliary functions \T{adjustCode'} and \T{adjust'} and modifying the definitions of \T{adjustCode} and \T{adjust} as follows:

\begin{small}
\begin{verbatim}
 OPNS
   adjustCode' : Bool -> Bit
   adjust' : Octet Octet Bool -> Octet
 RULES
   adjustCode (O) -> adjustCode' (needAdjust (O))

   adjustCode' (true)  -> x1
   adjustCode' (false) -> x0

   adjust (O, O') -> adjust' (O, O', needAdjust (O))

   adjust (O, O', true)  -> xorOctet (O, O')
   adjust (O, O', false) -> O
\end{verbatim}
\end{small}

\subsection{Definitions (2) of MAA-specific cryptographic functions}

We now define a second set of functions, namely the ``multiplicative'' functions used for MAA computations. The three principal operations are \T{MUL1}, \T{MUL2}, and \T{MUL2A}; the other ones are auxiliary functions.

\begin{small}
\begin{verbatim}
 OPNS
   MUL1 : Block Block -> Block
   MUL1XY : Pair -> Block
   MUL1UL : Block Block -> Block
   MUL1SC : Pair -> Block
   MUL2 : Block Block -> Block
   MUL2XY : Pair -> Block
   MUL2UL : Block Block -> Block
   MUL2DEL : Pair Block -> Block
   MUL2FL : Block Block -> Block
   MUL2SC : Pair -> Block
   MUL2A : Block Block -> Block
   MUL2AXY : Pair -> Block
   MUL2AUL : Block Block -> Block
   MUL2ADL : Block Block -> Block
   MUL2ASC : Pair -> Block
 VARS
   W W' Wcarry : Block
 RULES
   MUL1 (W, W') -> MUL1XY (MUL (W, W'))
   MUL1XY (buildPair (W, W')) -> MUL1UL (W, W')
   MUL1UL (W, W') -> MUL1SC (ADDC (W, W'))
   MUL1SC (buildPair (Wcarry, W)) -> ADD (W, Wcarry)

   MUL2 (W, W') -> MUL2XY (MUL (W, W'))
   MUL2XY (buildPair (W, W')) -> MUL2UL (W, W')
   MUL2UL (W, W') -> MUL2DEL (ADDC (W, W), W')
   MUL2DEL (buildPair (Wcarry, W), W') -> MUL2FL (ADD (W, ADD (Wcarry, Wcarry)), W')
   MUL2FL (W, W') -> MUL2SC (ADDC (W, W'))
   MUL2SC (buildPair (Wcarry, W)) -> ADD (W, ADD (Wcarry, Wcarry))

   MUL2A (W, W') -> MUL2AXY (MUL (W, W'))
   MUL2AXY (buildPair (W, W')) -> MUL2AUL (W, W')
   MUL2AUL (W, W') -> MUL2ADL (ADD (W, W), W')
   MUL2ADL (W, W') -> MUL2ASC (ADDC (W, W'))
   MUL2ASC (buildPair (Wcarry, W)) -> ADD (W, ADD (Wcarry, Wcarry))
\end{verbatim}
\end{small}

\subsection{Definitions (3) of MAA-specific cryptographic functions}

We now define a third set of functions used for MAA computations.

\begin{small}
\begin{verbatim}
 OPNS
   squareHalf : Half -> Block
   Q : Octet -> Block
   H4 : Block -> Block
   H6 : Block -> Block
   H8 : Block -> Block
   H0 : Block -> Block
   H5 : Block Octet -> Block
   H7 : Block -> Block
   H9 : Block  -> Block
   J1_2 : Block -> Block
   J1_4 : Block -> Block
   J1_6 : Block -> Block
   J1_8 : Block -> Block
   J2_2 : Block -> Block
   J2_4 : Block -> Block
   J2_6 : Block -> Block
   J2_8 : Block -> Block
   K1_2 : Block -> Block
   K1_4 : Block -> Block
   K1_5 : Block -> Block
   K1_7 : Block -> Block
   K1_9 : Block -> Block
   K2_2 : Block -> Block
   K2_4 : Block -> Block
   K2_5 : Block -> Block
   K2_7 : Block -> Block
   K2_9 : Block -> Block
 VARS
   H : Half
   O : Octet
   W : Block
 RULES
   squareHalf (H) -> mulHalf (H, H)

   Q (O) -> squareHalf (addHalf (buildHalf (x00, O), x0001))

   J1_2 (W) -> MUL1 (W, W)
   J1_4 (W) -> MUL1 (J1_2 (W), J1_2 (W))
   J1_6 (W) -> MUL1 (J1_2 (W), J1_4 (W))
   J1_8 (W) -> MUL1 (J1_2 (W), J1_6 (W))

   J2_2 (W) -> MUL2 (W, W)
   J2_4 (W) -> MUL2 (J2_2 (W), J2_2 (W))
   J2_6 (W) -> MUL2 (J2_2 (W), J2_4 (W))
   J2_8 (W) -> MUL2 (J2_2 (W), J2_6 (W))

   K1_2 (W) -> MUL1 (W, W)
   K1_4 (W) -> MUL1 (K1_2 (W), K1_2 (W))
   K1_5 (W) -> MUL1 (W, K1_4 (W))
   K1_7 (W) -> MUL1 (K1_2 (W), K1_5 (W))
   K1_9 (W) -> MUL1 (K1_2 (W), K1_7 (W))

   K2_2 (W) -> MUL2 (W, W)
   K2_4 (W) -> MUL2 (K2_2 (W), K2_2 (W))
   K2_5 (W) -> MUL2 (W, K2_4 (W))
   K2_7 (W) -> MUL2 (K2_2 (W), K2_5 (W))
   K2_9 (W) -> MUL2 (K2_2 (W), K2_7 (W))

   H4 (W) -> XOR (J1_4 (W), J2_4 (W))
   H6 (W) -> XOR (J1_6 (W), J2_6 (W))
   H8 (W) -> XOR (J1_8 (W), J2_8 (W))

   H0 (W) -> XOR (K1_5 (W), K2_5 (W))
   H5 (W, O) -> MUL2 (H0 (W), Q (O))
   H7 (W) -> XOR (K1_7 (W), K2_7 (W))
   H9 (W) -> XOR (K1_9 (W), K2_9 (W))
\end{verbatim}
\end{small}

\subsection{Definitions (4) of MAA-specific cryptographic functions}

We now define the higher-level functions that implement the MAA algorithm, namely the prelude, the inner loop, and the coda; the two principal functions are \T{MAA} (which computes the signature of a non-segmented message) and \T{MAC} (which splits a message into 1024-byte segments and computes the overall signature of this message by iterating on each segment, the 4-byte signature of each segment being prepended to the bytes of the next segment).

\begin{small}
\begin{verbatim}
 OPNS
   preludeXY : Block Block -> Pair
   preludeVW : Block Block -> Pair
   preludeST : Block Block -> Pair
   preludeXY' : Pair Octet -> Pair
   preludeVW' : Pair -> Pair
   preludeST' : Pair -> Pair
   computeXY : Pair Pair Block -> Pair
   computeXY' : Pair Block Block -> Pair
   computeVW : Pair -> Pair
   loop1 : Pair Pair Message -> Pair
   loop2 : Pair Pair Message -> Pair
   coda : Pair Pair Pair -> Block
   MAA : Key Message -> Block
   MAA' : Pair Pair Pair Message -> Block
   MAC : Key Message -> Block
   MACfirst : Key SegmentedMessage -> Block
   MACnext : Key Block SegmentedMessage -> Block
 VARS
   K : Key
   O : Block
   M : Message
   P P' P1 P2 P3 : Pair
   S : SegmentedMessage
   W W' W1 W2 : Block
 RULES
   % functions implementing the MAA prelude

   preludeXY (W1, W2) -> preludeXY' (BYT (W1, W2), PAT (W1, W2))
   preludeVW (W1, W2) -> preludeVW' (BYT (W1, W2))
   preludeST (W1, W2) -> preludeST' (BYT (W1, W2))

   preludeXY' (buildPair (W, W'), O) -> BYT (H4 (W), H5 (W', O))
   preludeVW' (buildPair (W, W'))    -> BYT (H6 (W), H7 (W'))
   preludeST' (buildPair (W, W'))    -> BYT (H8 (W), H9 (W'))

   % functions implementing the MAA inner loop

   computeXY (P, P', W) -> computeXY' (P, W, XOR' (computeVW (P')))

   computeXY' (buildPair (W1, W2), W, W')
   -> buildPair (MUL1 (XOR (W1, W), FIX1 (ADD (XOR (W2, W), W'))),
                 MUL2A (XOR (W2, W), FIX2 (ADD (XOR (W1, W), W'))))

   computeVW (buildPair (W1, W2)) -> buildPair (CYC (W1), W2)

   loop1 (P, P', unitMessage (W)) -> computeXY (P, P', W)
   loop1 (P, P', consMessage (W, M)) -> loop1 (computeXY (P, P', W), computeVW (P'), M)

   loop2 (P, P', unitMessage (W)) -> computeVW (P')
   loop2 (P, P', consMessage (W, M)) -> loop2 (computeXY (P, P', W), computeVW (P'), M)

   % function implementing the MAA coda

   coda (P, P', buildPair (W, W'))
   -> XOR' (computeXY (computeXY (P, P', W), computeVW (P'), W'))

   % functions computing the MAA on non-segmented messages

   MAA (buildKey (W1, W2), M)
   -> MAA' (preludeXY (W1, W2), preludeVW (W1, W2), preludeST (W1, W2), M)

   MAA' (P1, P2, P3, M) -> coda (loop1 (P1, P2, M), loop2 (P1, P2, M), P3)

   % functions computing the MAC on segmented messages

   MAC (K, M) -> MACfirst (K, splitSegment (M))

   MACfirst (K, unitSegment (M)) -> MAA (K, M)
   MACfirst (K, consSegment (M, S)) -> MACnext (K, MAA (K, M), S)

   MACnext (K, W, unitSegment (M)) -> MAA (K, consMessage (W, M))
   MACnext (K, W, consSegment (M, S)) -> MACnext (K, MAA (K, consMessage (W, M)), S)
\end{verbatim}
\end{small}

\subsection{Test vectors (1) for checking MAA computations}

We now define a first set of test vectors for the MAA. The following expressions implement the checks listed in Tables 1, 2, and~3 of \cite{Davies-Clayden-88} and should all evaluate to \T{true} if the MAA functions are correctly implemented.

\begin{small}
\begin{verbatim}
   % test vectors for function MUL1 - cf. Table 1
   eqBlock (MUL1 (x0000000F, x0000000E), x000000D2)
   eqBlock (MUL1 (xFFFFFFF0, x0000000E), xFFFFFF2D)
   eqBlock (MUL1 (xFFFFFFF0, xFFFFFFF1), x000000D2)

   % test vectors for function MUL2 - cf. Table 1
   eqBlock (MUL2 (x0000000F, x0000000E), x000000D2)
   eqBlock (MUL2 (xFFFFFFF0, x0000000E), xFFFFFF3A)
   eqBlock (MUL2 (xFFFFFFF0, xFFFFFFF1), x000000B6)

   % test vectors for function MUL2A - cf. Table 1
   eqBlock (MUL2A (x0000000F, x0000000E), x000000D2)
   eqBlock (MUL2A (xFFFFFFF0, x0000000E), xFFFFFF3A)
   eqBlock (MUL2A (x7FFFFFF0, xFFFFFFF1), x800000C2)
   eqBlock (MUL2A (xFFFFFFF0, x7FFFFFF1), x000000C4)

   % test vectors for function BYT - cf. Table 2
   eqPair (BYT (x00000000, x00000000), buildPair (x0103070F, x1F3F7FFF))
   eqPair (BYT (xFFFF00FF, xFFFFFFFF), buildPair (xFEFC07F0, xE0C08000))
   eqPair (BYT (xAB00FFCD, xFFEF0001), buildPair (xAB01FCCD, xF2EF3501))

   % test vectors for function PAT - cf. Table 2
   eqOctet (PAT (x00000000, x00000000), xFF)
   eqOctet (PAT (xFFFF00FF, xFFFFFFFF), xFF)
   eqOctet (PAT (xAB00FFCD, xFFEF0001), x6A)

   % test vectors for functions J1_i - cf. Table 3
   eqBlock (J1_2 (x00000100), x00010000)
   eqBlock (J1_4 (x00000100), x00000001)
   eqBlock (J1_6 (x00000100), x00010000)
   eqBlock (J1_8 (x00000100), x00000001)

   % test vectors for functions J2_i - cf. Table 3
   eqBlock (J2_2 (x00000100), x00010000)
   eqBlock (J2_4 (x00000100), x00000002)
   eqBlock (J2_6 (x00000100), x00020000)
   eqBlock (J2_8 (x00000100), x00000004)

   % test vectors for functions Hi - cf. Table 3
   eqBlock (H4 (x00000100), x00000003)
   eqBlock (H6 (x00000100), x00030000)
   eqBlock (H8 (x00000100), x00000005)

   % test vectors for functions K1_i - cf. Table 3
   eqBlock (K1_2 (x00000080), x00004000)
   eqBlock (K1_4 (x00000080), x10000000)
   eqBlock (K1_5 (x00000080), x00000008)
   eqBlock (K1_7 (x00000080), x00020000)
   eqBlock (K1_9 (x00000080), x80000000)

   % test vectors for functions K2_i - cf. Table 3
   eqBlock (K2_2 (x00000080), x00004000)
   eqBlock (K2_4 (x00000080), x10000000)
   eqBlock (K2_5 (x00000080), x00000010)
   eqBlock (K2_7 (x00000080), x00040000)
   eqBlock (K2_9 (x00000080), x00000002)

   % test vectors for functions Hi - cf. Table 3
   eqBlock (H0 (x00000080), x00000018)
   eqBlock (Q (x01), x00000004)
   eqBlock (H5 (x00000080, x01), x00000060)
   eqBlock (H7 (x00000080), x00060000)
   eqBlock (H9 (x00000080), x80000002)

   % test vectors for function PAT - cf. Table 3
   eqOctet (PAT (x00000003, x00000060), xEE)
   eqOctet (PAT (x00030000, x00060000), xBB)
   eqOctet (PAT (x00000005, x80000002), xE6)

   % test vectors for function BYT - inferred from Table 3
   eqPair (BYT (x00000003, x00000060), buildPair (x01030703, x1D3B7760)) % (X0, Y0)
   eqPair (BYT (x00030000, x00060000), buildPair (x0103050B, x17065DBB)) % (V0, W)
   eqPair (BYT (x00000005, x80000002), buildPair (x01030705, x80397302)) % (S, T)

\end{verbatim}
\end{small}

\subsection{Test vectors (2) for checking MAA computations}

We now define a second set of test vectors for the MAA, based upon Table~4 of \cite{Davies-Clayden-88}. The following expressions implement six groups of checks (three single-block messages and one three-block message). They should all evaluate to \T{true} if the main loop of MAA (as described page~10 of \cite{Davies-Clayden-88}) is correctly implemented.

\begin{small}
\begin{verbatim}
   % test vectors for the first single-block message
   eqBlock (CYC (x00000003), x00000006)                % V
   eqBlock (XOR (x00000006, x00000003), x00000005)     % E
   eqBlock (XOR (x00000002, x00000005), x00000007)     % X
   eqBlock (XOR (x00000003, x00000005), x00000006)     % Y
   eqBlock (ADD (x00000005, x00000006), x0000000B)     % F
   eqBlock (ADD (x00000005, x00000007), x0000000C)     % G
   eqBlock (OR (x0000000B, x00000004), x0000000F)      % F
   eqBlock (OR (x0000000C, x00000001), x0000000D)      % G
   eqBlock (AND (x0000000F, xFFFFFFF7), x00000007)     % F
   eqBlock (AND (x0000000D, xFFFFFFFB), x00000009)     % G
   eqBlock (MUL1 (x00000007, x00000007), x00000031)    % X
   eqBlock (MUL2A (x00000006, x00000009), x00000036)   % Y
   eqBlock (XOR (x00000031, x00000036), x00000007)     % Z

   % test vectors for the second single-block message
   eqBlock (CYC (x00000003), x00000006)                % V
   eqBlock (XOR (x00000006, x00000003), x00000005)     % E
   eqBlock (XOR (xFFFFFFFD, x00000001), xFFFFFFFC)     % X
   eqBlock (XOR (xFFFFFFFC, x00000001), xFFFFFFFD)     % Y
   eqBlock (ADD (x00000005, xFFFFFFFD), x00000002)     % F
   eqBlock (ADD (x00000005, xFFFFFFFC), x00000001)     % G
   eqBlock (OR (x00000002, x00000001), x00000003)      % F
   eqBlock (OR (x00000001, x00000004), x00000005)      % G
   eqBlock (AND (x00000003, xFFFFFFF9), x00000001)     % F
   eqBlock (AND (x00000005, xFFFFFFFC), x00000004)     % G
   eqBlock (MUL1 (xFFFFFFFC, x00000001), xFFFFFFFC)    % X
   eqBlock (MUL2A (xFFFFFFFD, x00000004), xFFFFFFFA)   % Y
   eqBlock (XOR (xFFFFFFFC, xFFFFFFFA), x00000006)     % Z

   % test vectors for the third single-block message
   eqBlock (CYC (x00000007), x0000000E)                % V
   eqBlock (XOR (x0000000E, x00000007), x00000009)     % E
   eqBlock (XOR (xFFFFFFFD, x00000008), xFFFFFFF5)     % X
   eqBlock (XOR (xFFFFFFFC, x00000008), xFFFFFFF4)     % Y
   eqBlock (ADD (x00000009, xFFFFFFF4), xFFFFFFFD)     % F
   eqBlock (ADD (x00000009, xFFFFFFF5), xFFFFFFFE)     % G
   eqBlock (OR (xFFFFFFFD, x00000001), xFFFFFFFD)      % F
   eqBlock (OR (xFFFFFFFE, x00000002), xFFFFFFFE)      % G
   eqBlock (AND (xFFFFFFFD, xFFFFFFFE), xFFFFFFFC)     % F
   eqBlock (AND (xFFFFFFFE, x7FFFFFFD), x7FFFFFFC)     % G
   eqBlock (MUL1 (xFFFFFFF5, xFFFFFFFC), x0000001E)    % X
   eqBlock (MUL2A (xFFFFFFF4, x7FFFFFFC), x0000001E)   % Y
   eqBlock (XOR (x0000001E, x0000001E), x00000000)     % Z

   % test vectors for three-block message: first block
   eqBlock (CYC (x00000001), x00000002)                % V
   eqBlock (XOR (x00000002, x00000001), x00000003)     % E
   eqBlock (XOR (x00000001, x00000000), x00000001)     % X
   eqBlock (XOR (x00000002, x00000000), x00000002)     % Y
   eqBlock (ADD (x00000003, x00000002), x00000005)     % F
   eqBlock (ADD (x00000003, x00000001), x00000004)     % G
   eqBlock (OR (x00000005, x00000002), x00000007)      % F
   eqBlock (OR (x00000004, x00000001), x00000005)      % G
   eqBlock (AND (x00000007, xFFFFFFFB), x00000003)     % F
   eqBlock (AND (x00000005, xFFFFFFFB), x00000001)     % G
   eqBlock (MUL1 (x00000001, x00000003), x00000003)    % X
   eqBlock (MUL2A (x00000002, x00000001), x00000002)   % Y
   eqBlock (XOR (x00000003, x00000002), x00000001)     % Z

   % test vectors for the three-block message: second block
   eqBlock (CYC (x00000002), x00000004)                % V
   eqBlock (XOR (x00000004, x00000001), x00000005)     % E
   eqBlock (XOR (x00000003, x00000001), x00000002)     % X
   eqBlock (XOR (x00000002, x00000001), x00000003)     % Y
   eqBlock (ADD (x00000005, x00000003), x00000008)     % F
   eqBlock (ADD (x00000005, x00000002), x00000007)     % G
   eqBlock (OR (x00000008, x00000002), x0000000A)      % F
   eqBlock (OR (x00000007, x00000001), x00000007)      % G
   eqBlock (AND (x0000000A, xFFFFFFFB), x0000000A)     % F
   eqBlock (AND (x00000007, xFFFFFFFB), x00000003)     % G
   eqBlock (MUL1 (x00000002, x0000000A), x00000014)    % X
   eqBlock (MUL2A (x00000003, x00000003), x00000009)   % Y
   eqBlock (XOR (x00000014, x00000009), x0000001D)     % Z

   % test vectors for three-block message: third block
   eqBlock (CYC (x00000004), x00000008)                % V
   eqBlock (XOR (x00000008, x00000001), x00000009)     % E
   eqBlock (XOR (x00000014, x00000002), x00000016)     % X
   eqBlock (XOR (x00000009, x00000002), x0000000B)     % Y
   eqBlock (ADD (x00000009, x0000000B), x00000014)     % F
   eqBlock (ADD (x00000009, x00000016), x0000001F)     % G
   eqBlock (OR (x00000014, x00000002), x00000016)      % F
   eqBlock (OR (x0000001F, x00000001), x0000001F)      % G
   eqBlock (AND (x00000016, xFFFFFFFB), x00000012)     % F
   eqBlock (AND (x0000001F, xFFFFFFFB), x0000001B)     % G
   eqBlock (MUL1 (x00000016, x00000012), x0000018C)    % X
   eqBlock (MUL2A (x0000000B, x0000001B), x00000129)   % Y
   eqBlock (XOR (x0000018C, x00000129), x000000A5)     % Z
\end{verbatim}
\end{small}

\noindent We complete the above tests with additional test vectors taken from \cite[Annex~E.3.3]{ISO-8730:1990}, which only gives detailed values for the first block of the 84-block test message.

\begin{small}
\begin{verbatim}
   % test vectors for block x0A202020 with key (J = xE6A12F07, K = x9D15C437)
   eqBlock (CYC (xC4EB1AEB), x89D635D7)                % V
   eqBlock (XOR (x89D635D7, xF6A09667), x7F76A3B0)     % E
   eqBlock (XOR (x21D869BA, x0A202020), x2BF8499A)     % X
   eqBlock (XOR (x7792F9D4, x0A202020), x7DB2D9F4)     % Y
   eqBlock (ADD (x7F76A3B0, x7DB2D9F4), xFD297DA4)     % F
   eqBlock (ADD (x7F76A3B0, x2BF8499A), xAB6EED4A)     % G
   eqBlock (OR (xFD297DA4, x02040801), xFF2D7DA5)      % F
   eqBlock (OR (xAB6EED4A, x00804021), xABEEED6B)      % G
   eqBlock (AND (xFF2D7DA5, xBFEF7FDF), xBF2D7D85)     % F
   eqBlock (AND (xABEEED6B, x7DFEFBFF), x29EEE96B)     % G
   eqBlock (MUL1 (x2BF8499A, xBF2D7D85), x0AD67E20)    % X
   eqBlock (MUL2A (x7DB2D9F4, x29EEE96B), x30261492)   % Y
\end{verbatim}
\end{small}

\subsection{Test vectors (3) for checking MAA computations}

We now define a third set of test vectors for the MAA, based upon Table~5 of \cite{Davies-Clayden-88}. The following expressions implement four groups of checks, with two different keys and two different messages. They should all evaluate to \T{true} if the MAA signature is correctly computed.

\begin{small}
\begin{verbatim}
   % test vectors of the first column of Table 5
   % key (J = x00FF00FF, K = x00000000), message (M1 = x55555555, M2 = xAAAAAAAA)

   eqOctet (PAT (x00FF00FF, x00000000), xFF)                          % P
   eqPair (preludeXY (x00FF00FF, x00000000),
           buildPair (x4A645A01, x50DEC930))                          % (X0, Y0)
   eqPair (preludeVW (x00FF00FF, x00000000),
           buildPair (x5CCA3239, xFECCAA6E))                          % (VO, W)
   eqPair (preludeST (x00FF00FF, x00000000),
           buildPair (x51EDE9C7, x24B66FB5))                          % (S, T)

   eqPair (computeXY' (buildPair (x4A645A01, x50DEC930), x55555555,
           XOR (nCYC (n1, x5CCA3239), xFECCAA6E)),                    % 1st iteration
           buildPair (x48B204D6, x5834A585))                          % (X, Y)
   eqPair (computeXY' (buildPair (x48B204D6, x5834A585), xAAAAAAAA,
           XOR (nCYC (n2, x5CCA3239), xFECCAA6E)),                    % 2nd iteration
           buildPair (x4F998E01, xBE9F0917))                          % (X, Y)
   eqPair (computeXY' (buildPair (x4F998E01, xBE9F0917), x51EDE9C7,
           XOR (nCYC (n3, x5CCA3239), xFECCAA6E)),                    % coda: use of S
           buildPair (x344925FC, xDB9102B0))                          % (X, Y)
   eqPair (computeXY' (buildPair (x344925FC, xDB9102B0), x24B66FB5,
           XOR (nCYC (n4, x5CCA3239), xFECCAA6E)),                    % coda: use of T
           buildPair (x277B4B25, xD636250D))                          % (X, Y)

   eqBlock (XOR (x277B4B25, xD636250D), xF14D6E28)                    % Z (i.e., MAA)

   % test vectors of the second column of Table 5
   % key (J = x00FF00FF, K = x00000000), message (M1 = x55555555, M2 = xAAAAAAAA)

   eqOctet (PAT (x00FF00FF, x00000000), xFF)                          % P
   eqPair (preludeXY (x00FF00FF, x00000000),
           buildPair (x4A645A01, x50DEC930))                          % (X0, Y0)
   eqPair (preludeVW (x00FF00FF, x00000000),
           buildPair (x5CCA3239, xFECCAA6E))                          % (VO, W)
   eqPair (preludeST (x00FF00FF, x00000000),
           buildPair (x51EDE9C7, x24B66FB5))                          % (S, T)

   eqPair (computeXY' (buildPair (x4A645A01, x50DEC930), xAAAAAAAA,
           XOR (nCYC (n1, x5CCA3239), xFECCAA6E)),                    % 1st iteration
           buildPair (x6AEBACF8, x9DB15CF6))                          % (X, Y)
   eqPair (computeXY' (buildPair (x6AEBACF8, x9DB15CF6), x55555555,
           XOR (nCYC (n2, x5CCA3239), xFECCAA6E)),                    % 2nd iteration
           buildPair (x270EEDAF, xB8142629))                          % (X, Y)
   eqPair (computeXY' (buildPair (x270EEDAF, xB8142629), x51EDE9C7,
           XOR (nCYC (n3, x5CCA3239), xFECCAA6E)),                    % coda: use of S
           buildPair (x29907CD8, xBA92DB12))                          % (X, Y)
   eqPair (computeXY' (buildPair (x29907CD8, xBA92DB12), x24B66FB5,
           XOR (nCYC (n4, x5CCA3239), xFECCAA6E)),                    % coda: use of T
           buildPair (x28EAD8B3, x81D10CA3))                          % (X, Y)

   eqBlock (XOR (x28EAD8B3, x81D10CA3), xA93BD410)                    % Z (i.e., MAA)

   % test vectors of the third column of Table 5
   % key (J = x55555555, K = x5A35D667), message (M1 = x00000000, M2 = xFFFFFFFF)

   eqOctet (PAT (x55555555, x5A35D667), x00)                          % P
   eqPair (preludeXY (x55555555, x5A35D667),
           buildPair (x34ACF886, x7397C9AE))                          % (X0, Y0)
   eqPair (preludeVW (x55555555, x5A35D667),
           buildPair (x7201F4DC, x2829040B))                          % (VO, W)
   eqPair (preludeST (x55555555, x5A35D667),
           buildPair (x9E2E7B36, x13647149))                          % (S, T)

   eqPair (computeXY' (buildPair (x34ACF886, x7397C9AE), x00000000,
           XOR (nCYC (n1, x7201F4DC), x2829040B)),                    % 1st iteration
           buildPair (x2FD76FFB, x550D91CE))                          % (X, Y)
   eqPair (computeXY' (buildPair (x2FD76FFB, x550D91CE), xFFFFFFFF,
           XOR (nCYC (n2, x7201F4DC), x2829040B)),                    % 2nd iteration
           buildPair (xA70FC148, x1D10D8D3))                          % (X, Y)
   eqPair (computeXY' (buildPair (xA70FC148, x1D10D8D3), x9E2E7B36,
           XOR (nCYC (n3, x7201F4DC), x2829040B)),                    % coda: use of S
           buildPair (xB1CC1CC5, x29C1485F))                          % (X, Y)
   eqPair (computeXY' (buildPair (xB1CC1CC5, x29C1485F), x13647149,
           XOR (nCYC (n4, x7201F4DC), x2829040B)),                    % coda: use of T
           buildPair (x288FC786, x9115A558))                          % (X, Y)

   eqBlock (XOR (x288FC786, x9115A558), xB99A62DE)                    % Z (i.e., MAA)

   % test vectors of the fourth column of Table 5
   % key (J = x55555555, K = x5A35D667), message (M1 = xFFFFFFFF, M2 = x00000000)

   eqOctet (PAT (x55555555, x5A35D667), x00)                          % P
   eqPair (preludeXY (x55555555, x5A35D667),
           buildPair (x34ACF886, x7397C9AE))                          % (X0, Y0)
   eqPair (preludeVW (x55555555, x5A35D667),
           buildPair (x7201F4DC, x2829040B))                          % (VO, W)
   eqPair (preludeST (x55555555, x5A35D667),
           buildPair (x9E2E7B36, x13647149))                          % (S, T)

   eqPair (computeXY' (buildPair (x34ACF886, x7397C9AE), xFFFFFFFF,
           XOR (nCYC (n1, x7201F4DC), x2829040B)),                    % 1st iteration
           buildPair (x8DC8BBDE, xFE4E5BDD))                          % (X, Y)
   eqPair (computeXY' (buildPair (x8DC8BBDE, xFE4E5BDD), x00000000,
           XOR (nCYC (n2, x7201F4DC), x2829040B)),                    % 2nd iteration
           buildPair (xCBC865BA, x0297AF6F))                          % (X, Y)
   eqPair (computeXY' (buildPair (xCBC865BA, x0297AF6F), x9E2E7B36,
           XOR (nCYC (n3, x7201F4DC), x2829040B)),                    % coda: use of S
           buildPair (x3CF3A7D2, x160EE9B5))                          % (X, Y)
   eqPair (computeXY' (buildPair (x3CF3A7D2, x160EE9B5), x13647149,
           XOR (nCYC (n4, x7201F4DC), x2829040B)),                    % coda: use of T
           buildPair (xD0482465, x7050EC5E))                          % (X, Y)

   eqBlock (XOR (xD0482465, x7050EC5E), xA018C83B)                    % Z (i.e., MAA)
\end{verbatim}
\end{small}

\noindent We complete the above tests with additional test vectors taken from \cite[Annex~E.3.3]{ISO-8730:1990}, which gives prelude results computed for another key.

\begin{small}
\begin{verbatim}
   % key (J = xE6A12F07, K = x9D15C437)
   eqPair (preludeXY (xE6A12F07, x9D15C437),
           buildPair (x21D869BA, x7792F9D4))                          % (X0, Y0)
   eqPair (preludeVW (xE6A12F07, x9D15C437),
           buildPair (xC4EB1AEB, xF6A09667))                          % (VO, W)
   eqPair (preludeST (xE6A12F07, x9D15C437),
           buildPair (x6D67E884, xA511987A))                          % (S, T)
\end{verbatim}
\end{small}

\subsection{Test vectors (4) for checking MAA computations}

	We define a last set of test vectors for the MAA. The first one (a message of 20 blocks containing only zeros) was directly taken from Table~6 of \cite{Davies-Clayden-88}.

\begin{small}
\begin{verbatim}
   eqPair (computeXY' (buildPair (x204E80A7, x077788A2), x00000000,
           XOR (nCYC (n1, x17A808FD), xFEA1D334)),                    % 1st iteration
           buildPair (x303FF4AA, x1277A6D4))                          % (X, Y)

   eqPair (computeXY' (buildPair (x303FF4AA, x1277A6D4), x00000000,
           XOR (nCYC (n2, x17A808FD), xFEA1D334)),                    % 2nd iteration
           buildPair (x55DD063F, x4C49AAE0))                          % (X, Y)

   eqPair (computeXY' (buildPair (x55DD063F, x4C49AAE0), x00000000,
           XOR (nCYC (n3, x17A808FD), xFEA1D334)),                    % 3rd iteration
           buildPair (x51AF3C1D, x5BC02502))                          % (X, Y)

   eqPair (computeXY' (buildPair (x51AF3C1D, x5BC02502), x00000000,
           XOR (nCYC (n4, x17A808FD), xFEA1D334)),                    % 4th iteration
           buildPair (xA44AAAC0, x63C70DBA))                          % (X, Y)

   eqPair (computeXY' (buildPair (xA44AAAC0, x63C70DBA), x00000000,
           XOR (nCYC (n5, x17A808FD), xFEA1D334)),                    % 5th iteration
           buildPair (x4D53901A, x2E80AC30))                          % (X, Y)

   eqPair (computeXY' (buildPair (x4D53901A, x2E80AC30), x00000000,
           XOR (nCYC (n6, x17A808FD), xFEA1D334)),                    % 6th iteration
           buildPair (x5F38EEF1, x2A6091AE))                          % (X, Y)

   eqPair (computeXY' (buildPair (x5F38EEF1, x2A6091AE), x00000000,
           XOR (nCYC (n7, x17A808FD), xFEA1D334)),                    % 7th iteration
           buildPair (xF0239DD5, x3DD81AC6))                          % (X, Y)

   eqPair (computeXY' (buildPair (xF0239DD5, x3DD81AC6), x00000000,
           XOR (nCYC (n8, x17A808FD), xFEA1D334)),                    % 8th iteration
           buildPair (xEB35B97F, x9372CDC6))                          % (X, Y)

   eqPair (computeXY' (buildPair (xEB35B97F, x9372CDC6), x00000000,
           XOR (nCYC (n9, x17A808FD), xFEA1D334)),                    % 9th iteration
           buildPair (x4DA124A1, xC6B1317E))                          % (X, Y)

   eqPair (computeXY' (buildPair (x4DA124A1, xC6B1317E), x00000000,
           XOR (nCYC (n10, x17A808FD), xFEA1D334)),                   % 10th iteration
           buildPair (x7F839576, x74B39176))                          % (X, Y)

   eqPair (computeXY' (buildPair (x7F839576, x74B39176), x00000000,
           XOR (nCYC (n11, x17A808FD), xFEA1D334)),                   % 11th iteration
           buildPair (x11A9D254, xD78634BC))                          % (X, Y)

   eqPair (computeXY' (buildPair (x11A9D254, xD78634BC), x00000000,
           XOR (nCYC (n12, x17A808FD), xFEA1D334)),                   % 12th iteration
           buildPair (xD8804CA5, xFDC1A8BA))                          % (X, Y)

   eqPair (computeXY' (buildPair (xD8804CA5, xFDC1A8BA), x00000000,
           XOR (nCYC (n13, x17A808FD), xFEA1D334)),                   % 13th iteration
           buildPair (x3F6F7248, x11AC46B8))                          % (X, Y)

   eqPair (computeXY' (buildPair (x3F6F7248, x11AC46B8), x00000000,
           XOR (nCYC (n14, x17A808FD), xFEA1D334)),                   % 14th iteration
           buildPair (xACBC13DD, x33D5A466))                          % (X, Y)

   eqPair (computeXY' (buildPair (xACBC13DD, x33D5A466), x00000000,
           XOR (nCYC (n15, x17A808FD), xFEA1D334)),                   % 15th iteration
           buildPair (x4CE933E1, xC21A1846))                          % (X, Y)

   eqPair (computeXY' (buildPair (x4CE933E1, xC21A1846), x00000000,
           XOR (nCYC (n16, x17A808FD), xFEA1D334)),                   % 16th iteration
           buildPair (xC1ED90DD, xCD959B46))                          % (X, Y)

   eqPair (computeXY' (buildPair (xC1ED90DD, xCD959B46), x00000000,
           XOR (nCYC (n17, x17A808FD), xFEA1D334)),                   % 17th iteration
           buildPair (x3CD54DEB, x613F8E2A))                          % (X, Y)

   eqPair (computeXY' (buildPair (x3CD54DEB, x613F8E2A), x00000000,
           XOR (nCYC (n18, x17A808FD), xFEA1D334)),                   % 18th iteration
           buildPair (xBBA57835, x07C72EAA))                          % (X, Y)

   eqPair (computeXY' (buildPair (xBBA57835, x07C72EAA), x00000000,
           XOR (nCYC (n19, x17A808FD), xFEA1D334)),                   % 19th iteration
           buildPair (xD7843FDC, x6AD6E8A4))                          % (X, Y)

   eqPair (computeXY' (buildPair (xD7843FDC, x6AD6E8A4), x00000000,
           XOR (nCYC (n20, x17A808FD), xFEA1D334)),                   % 20th iteration
           buildPair (x5EBA06C2, x91896CFA))                          % (X, Y)

   eqPair (computeXY' (buildPair (x5EBA06C2, x91896CFA), x76232E5F,
           XOR (nCYC (n21, x17A808FD), xFEA1D334)),                   % coda: use of S
           buildPair (x1D9C9655, x98D1CC75))                          % (X, Y)

   eqPair (computeXY' (buildPair (x1D9C9655, x98D1CC75), x4FB1138A,
           XOR (nCYC (n22, x17A808FD), xFEA1D334)),                   % coda: use of T
           buildPair (x7BC180AB, xA0B87B77))                          % (X, Y)

   eqBlock (MAC (buildKey (x80018001, x80018000),
                 makeMessage (n20, x00000000, x00000000)), xDB79FBDC)
\end{verbatim}
\end{small}

\noindent We believe that the test vector above is not sufficient to detect implementation mistakes arising from byte permutations (e.g., endianness issues) or incorrect segmentation of messages longer than 1024 bytes (i.e., 256 blocks). To address these issues, we added three supplementary test vectors that operate on messages of 16, 256, and 4100 blocks containing bit patterns not preserved by permutations.

\begin{small}
\begin{verbatim}
   eqBlock (MAC (buildKey (x80018001, x80018000),
                 makeMessage (n16, x00000000, x07050301)), x8CE37709)

   eqBlock (MAC (buildKey (x80018001, x80018000),
                 makeMessage (n256, x00000000, x07050301)), x717153D5)

   eqBlock (MAC (buildKey (x80018001, x80018000),
                 makeMessage (n4100, x00000000, x07050301)), x7783C51D)
\end{verbatim}
\end{small}

\subsection{Possible variants}

	The REC specification given in the present Annex could be enhanced in two directions that diverge from the modelling choices originally done in \cite{Munster-91-a} and could be given as exercises to students:

\begin{itemize}
	\item At present, the \T{Prelude} function is called several times when computing the MAC value for a given message; precisely, this function is called for every 256-block segment of the message. This is neither useful nor efficient. Propose a modification of the REC specification to ensure that the \T{Prelude} function is called only once per message.

	\item Before computing the MAC value for a given message, the REC specification converts this message into a segmented message by calling the \T{splitSegment} function. Actually, such a preliminary duplication of message contents is not mandatory and could be avoided. Propose a modification of the REC specification in which the \T{SegmentedMessage} sort and all the definitions of Section~\ref{SEGMENTS} are removed, so that the MAC value is directly computed using a one-pass traversal of the message list, from its head to its tail, still taking the MAA ``mode of operation'' into account.
\end{itemize}


\begin{thebibliography}{10}
\providecommand{\bibitemdeclare}[2]{}
\providecommand{\surnamestart}{}
\providecommand{\surnameend}{}
\providecommand{\urlprefix}{Available at }
\providecommand{\url}[1]{\texttt{#1}}
\providecommand{\href}[2]{\texttt{#2}}
\providecommand{\urlalt}[2]{\href{#1}{#2}}
\providecommand{\doi}[1]{doi:\urlalt{http://dx.doi.org/#1}{#1}}
\providecommand{\bibinfo}[2]{#2}

\bibitemdeclare{inproceedings}{Davies-85}
\bibitem{Davies-85}
\bibinfo{author}{Donald~W. \surnamestart Davies\surnameend}
  (\bibinfo{year}{1985}): \emph{\bibinfo{title}{{A Message Authenticator
  Algorithm Suitable for A Mainframe Computer}}}.
\newblock In \bibinfo{editor}{G.~R. \surnamestart Blakley\surnameend} \&
  \bibinfo{editor}{David \surnamestart Chaum\surnameend}, editors: {\sl
  \bibinfo{booktitle}{Advances in Cryptology -- Proceedings of the Workshop on
  the Theory and Application of Cryptographic Techniques (CRYPTO'84), Santa
  Barbara, CA, USA}}, {\sl \bibinfo{series}{Lecture Notes in Computer Science}}
  \bibinfo{volume}{196}, \bibinfo{publisher}{Springer}, pp.
  \bibinfo{pages}{393--400}, \doi{10.1007/3-540-39568-7_30}.

\bibitemdeclare{techreport}{Davies-Clayden-88}
\bibitem{Davies-Clayden-88}
\bibinfo{author}{Donald~W. \surnamestart Davies\surnameend} \&
  \bibinfo{author}{David~O. \surnamestart Clayden\surnameend}
  (\bibinfo{year}{1988}): \emph{\bibinfo{title}{{The Message Authenticator
  Algorithm (MAA) and its Implementation}}}.
\newblock \bibinfo{type}{NPL Report DITC} \bibinfo{number}{109/88},
  \bibinfo{institution}{National Physical Laboratory},
  \bibinfo{address}{Teddington, Middlesex, UK}.
\newblock \urlprefix\url{http://www.cix.co.uk/~klockstone/maa.pdf}.

\bibitemdeclare{inproceedings}{Didrich-Fett-Gerke-et-al-94}
\bibitem{Didrich-Fett-Gerke-et-al-94}
\bibinfo{author}{Klaus \surnamestart Didrich\surnameend},
  \bibinfo{author}{Andreas \surnamestart Fett\surnameend},
  \bibinfo{author}{Carola \surnamestart Gerke\surnameend},
  \bibinfo{author}{Wolfgang \surnamestart Grieskamp\surnameend} \&
  \bibinfo{author}{Peter \surnamestart Pepper\surnameend}
  (\bibinfo{year}{1994}): \emph{\bibinfo{title}{{OPAL: Design and
  Implementation of an Algebraic Programming Language}}}.
\newblock In \bibinfo{editor}{J{\"{u}}rg \surnamestart Gutknecht\surnameend},
  editor: {\sl \bibinfo{booktitle}{Proceedings of the International Conference
  on Programming Languages and System Architectures, Zurich, Switzerland}},
  {\sl \bibinfo{series}{Lecture Notes in Computer Science}}
  \bibinfo{volume}{782}, \bibinfo{publisher}{Springer}, pp.
  \bibinfo{pages}{228--244}, \doi{10.1007/3-540-57840-4_34}.

\bibitemdeclare{inproceedings}{Duran-Roldan-Bach-10}
\bibitem{Duran-Roldan-Bach-10}
\bibinfo{author}{Francisco \surnamestart Dur{\'{a}}n\surnameend},
  \bibinfo{author}{Manuel \surnamestart Rold{\'{a}}n\surnameend},
  \bibinfo{author}{Jean{-}Christophe \surnamestart Bach\surnameend},
  \bibinfo{author}{Emilie \surnamestart Balland\surnameend},
  \bibinfo{author}{Mark \surnamestart van~den Brand\surnameend},
  \bibinfo{author}{James~R. \surnamestart Cordy\surnameend},
  \bibinfo{author}{Steven \surnamestart Eker\surnameend}, \bibinfo{author}{Luc
  \surnamestart Engelen\surnameend}, \bibinfo{author}{Maartje \surnamestart
  de~Jonge\surnameend}, \bibinfo{author}{Karl~Trygve \surnamestart
  Kalleberg\surnameend}, \bibinfo{author}{Lennart C.~L. \surnamestart
  Kats\surnameend}, \bibinfo{author}{Pierre-Etienne \surnamestart
  Moreau\surnameend} \& \bibinfo{author}{Eelco \surnamestart Visser\surnameend}
  (\bibinfo{year}{2010}): \emph{\bibinfo{title}{{The Third Rewrite Engines
  Competition}}}.
\newblock In \bibinfo{editor}{Peter~Csaba \surnamestart
  {\"{O}}lveczky\surnameend}, editor: {\sl \bibinfo{booktitle}{Proceedings of
  the 8th International Workshop on Rewriting Logic and Its Applications
  (WRLA'10), Paphos, Cyprus}}, {\sl \bibinfo{series}{Lecture Notes in Computer
  Science}} \bibinfo{volume}{6381}, \bibinfo{publisher}{Springer}, pp.
  \bibinfo{pages}{243--261}, \doi{10.1007/978-3-642-16310-4_16}.

\bibitemdeclare{article}{Duran-Roldan-Balland-et-al-09}
\bibitem{Duran-Roldan-Balland-et-al-09}
\bibinfo{author}{Francisco \surnamestart Dur{\'{a}}n\surnameend},
  \bibinfo{author}{Manuel \surnamestart Rold{\'{a}}n\surnameend},
  \bibinfo{author}{Emilie \surnamestart Balland\surnameend},
  \bibinfo{author}{Mark \surnamestart van~den Brand\surnameend},
  \bibinfo{author}{Steven \surnamestart Eker\surnameend},
  \bibinfo{author}{Karl~Trygve \surnamestart Kalleberg\surnameend},
  \bibinfo{author}{Lennart C.~L. \surnamestart Kats\surnameend},
  \bibinfo{author}{Pierre-Etienne \surnamestart Moreau\surnameend},
  \bibinfo{author}{Ruslan \surnamestart Schevchenko\surnameend} \&
  \bibinfo{author}{Eelco \surnamestart Visser\surnameend}
  (\bibinfo{year}{2009}): \emph{\bibinfo{title}{{The Second Rewrite Engines
  Competition}}}.
\newblock {\sl \bibinfo{journal}{Electronic Notes in Theoretical Computer
  Science}} \bibinfo{volume}{238}(\bibinfo{number}{3}), pp.
  \bibinfo{pages}{281--291}, \doi{10.1016/j.entcs.2009.05.025}.

\bibitemdeclare{inproceedings}{Garavel-89-c}
\bibitem{Garavel-89-c}
\bibinfo{author}{Hubert \surnamestart Garavel\surnameend}
  (\bibinfo{year}{1989}): \emph{\bibinfo{title}{Compilation of LOTOS Abstract
  Data Types}}.
\newblock In \bibinfo{editor}{Son~T. \surnamestart Vuong\surnameend}, editor:
  {\sl \bibinfo{booktitle}{Proceedings of the 2nd International Conference on
  Formal Description Techniques {FORTE}'89 (Vancouver B.C., Canada)}},
  \bibinfo{publisher}{North-Holland}, pp. \bibinfo{pages}{147--162}.
\newblock \urlprefix\url{http://cadp.inria.fr/publications/Garavel-89-c.html}.

\bibitemdeclare{inproceedings}{Garavel-Turlier-93}
\bibitem{Garavel-Turlier-93}
\bibinfo{author}{Hubert \surnamestart Garavel\surnameend} \&
  \bibinfo{author}{Philippe \surnamestart Turlier\surnameend}
  (\bibinfo{year}{1993}): \emph{\bibinfo{title}{{C{\AE}SAR.ADT~: un compilateur
  pour les types abstraits alg\'ebriques du langage LOTOS}}}.
\newblock In \bibinfo{editor}{Rachida \surnamestart Dssouli\surnameend} \&
  \bibinfo{editor}{Gregor \surnamestart v.~Bochmann\surnameend}, editors: {\sl
  \bibinfo{booktitle}{Actes du Colloque Francophone pour l'Ing\'enierie des
  Protocoles (CFIP'93), Montr\'eal, Canada}}.
\newblock
  \urlprefix\url{http://cadp.inria.fr/publications/Garavel-Turlier-93.html}.

\bibitemdeclare{inproceedings}{Giesl-Brockschmidt-Emmes-et-al-14}
\bibitem{Giesl-Brockschmidt-Emmes-et-al-14}
\bibinfo{author}{J\"{u}rgen \surnamestart Giesl\surnameend},
  \bibinfo{author}{Marc \surnamestart Brockschmidt\surnameend},
  \bibinfo{author}{Fabian \surnamestart Emmes\surnameend},
  \bibinfo{author}{Florian \surnamestart Frohn\surnameend},
  \bibinfo{author}{Carsten \surnamestart Fuhs\surnameend},
  \bibinfo{author}{Carsten \surnamestart Otto\surnameend},
  \bibinfo{author}{Martin \surnamestart Pl\"{u}cker\surnameend},
  \bibinfo{author}{Peter \surnamestart Schneider{-}Kamp\surnameend},
  \bibinfo{author}{Thomas \surnamestart Str\"{o}der\surnameend},
  \bibinfo{author}{Stephanie \surnamestart Swiderski\surnameend} \&
  \bibinfo{author}{Ren{\'{e}} \surnamestart Thiemann\surnameend}
  (\bibinfo{year}{2014}): \emph{\bibinfo{title}{{Proving Termination of
  Programs Automatically with AProVE}}}.
\newblock In \bibinfo{editor}{St\'{e}phane \surnamestart Demri\surnameend},
  \bibinfo{editor}{Deepak \surnamestart Kapur\surnameend} \&
  \bibinfo{editor}{Christoph \surnamestart Weidenbach\surnameend}, editors:
  {\sl \bibinfo{booktitle}{Proceedings of the 7th International Joint
  Conference on Automated Reasoning (IJCAR'14), Vienna, Austria}}, {\sl
  \bibinfo{series}{Lecture Notes in Computer Science}} \bibinfo{volume}{8562},
  \bibinfo{publisher}{Springer}, pp. \bibinfo{pages}{184--191},
  \doi{10.1007/978-3-319-08587-6_13}.
\newblock
  \urlprefix\url{http://verify.rwth-aachen.de/giesl/papers/IJCAR14-AProVE.pdf}.

\bibitemdeclare{techreport}{ISO-8731-2:1987}
\bibitem{ISO-8731-2:1987}
\bibinfo{author}{\surnamestart ISO\surnameend} (\bibinfo{year}{1987}):
  \emph{\bibinfo{title}{{Approved Algorithms for Message Authentication -- Part
  2: Message Authenticator Algorithm (MAA)}}}.
\newblock \bibinfo{type}{International Standard} \bibinfo{number}{8731-2},
  \bibinfo{institution}{International Organization for Standardization --
  Banking}, \bibinfo{address}{Geneva}.

\bibitemdeclare{techreport}{ISO-8730:1990}
\bibitem{ISO-8730:1990}
\bibinfo{author}{\surnamestart ISO\surnameend} (\bibinfo{year}{1990}):
  \emph{\bibinfo{title}{{Requirements for Message Authentication
  (Wholesale)}}}.
\newblock \bibinfo{type}{International Standard} \bibinfo{number}{8730},
  \bibinfo{institution}{International Organization for Standardization --
  Banking}, \bibinfo{address}{Geneva}.

\bibitemdeclare{techreport}{ISO-8731-2:1992}
\bibitem{ISO-8731-2:1992}
\bibinfo{author}{\surnamestart ISO\surnameend} (\bibinfo{year}{1992}):
  \emph{\bibinfo{title}{{Approved Algorithms for Message Authentication -- Part
  2: Message Authenticator Algorithm}}}.
\newblock \bibinfo{type}{International Standard} \bibinfo{number}{8731-2},
  \bibinfo{institution}{International Organization for Standardization --
  Banking}, \bibinfo{address}{Geneva}.

\bibitemdeclare{techreport}{Lai-91}
\bibitem{Lai-91}
\bibinfo{author}{{M. K. F.} \surnamestart Lai\surnameend}
  (\bibinfo{year}{1991}): \emph{\bibinfo{title}{{A Formal Interpretation of the
  MAA Standard in Z}}}.
\newblock \bibinfo{type}{NPL Report DITC} \bibinfo{number}{184/91},
  \bibinfo{institution}{National Physical Laboratory},
  \bibinfo{address}{Teddington, Middlesex, UK}.

\bibitemdeclare{techreport}{Lampard-91}
\bibitem{Lampard-91}
\bibinfo{author}{R.~P. \surnamestart Lampard\surnameend}
  (\bibinfo{year}{1991}): \emph{\bibinfo{title}{{An Implementation of MAA from
  a VDM Specification}}}.
\newblock \bibinfo{type}{NPL Technical Memorandum DITC}
  \bibinfo{number}{50/91}, \bibinfo{institution}{National Physical Laboratory},
  \bibinfo{address}{Teddington, Middlesex, UK}.

\bibitemdeclare{book}{Menezes-vanOorschot-Vanstone-96}
\bibitem{Menezes-vanOorschot-Vanstone-96}
\bibinfo{author}{Alfred \surnamestart Menezes\surnameend},
  \bibinfo{author}{Paul~C. \surnamestart van Oorschot\surnameend} \&
  \bibinfo{author}{Scott~A. \surnamestart Vanstone\surnameend}
  (\bibinfo{year}{1996}): \emph{\bibinfo{title}{{Handbook of Applied
  Cryptography}}}.
\newblock \bibinfo{publisher}{CRC Press}, \doi{10.1201/9781439821916}.
\newblock \urlprefix\url{http://cacr.uwaterloo.ca/hac}.

\bibitemdeclare{techreport}{Munster-91-a}
\bibitem{Munster-91-a}
\bibinfo{author}{Harold~B. \surnamestart Munster\surnameend}
  (\bibinfo{year}{1991}): \emph{\bibinfo{title}{{LOTOS Specification of the MAA
  Standard, with an Evaluation of LOTOS}}}.
\newblock \bibinfo{type}{NPL Report DITC} \bibinfo{number}{191/91},
  \bibinfo{institution}{National Physical Laboratory},
  \bibinfo{address}{Teddington, Middlesex, UK}.
\newblock
  \urlprefix\url{ftp://ftp.inrialpes.fr/pub/vasy/publications/others/Munster-9%
1-a.pdf}.

\bibitemdeclare{techreport}{Parkin-ONeill-90}
\bibitem{Parkin-ONeill-90}
\bibinfo{author}{Graeme~I. \surnamestart Parkin\surnameend} \&
  \bibinfo{author}{G.~\surnamestart O'Neill\surnameend} (\bibinfo{year}{1990}):
  \emph{\bibinfo{title}{{Specification of the MAA Standard in VDM}}}.
\newblock \bibinfo{type}{NPL Report DITC} \bibinfo{number}{160/90},
  \bibinfo{institution}{National Physical Laboratory},
  \bibinfo{address}{Teddington, Middlesex, UK}.

\bibitemdeclare{inproceedings}{Parkin-ONeill-91}
\bibitem{Parkin-ONeill-91}
\bibinfo{author}{Graeme~I. \surnamestart Parkin\surnameend} \&
  \bibinfo{author}{G.~\surnamestart O'Neill\surnameend} (\bibinfo{year}{1991}):
  \emph{\bibinfo{title}{{Specification of the MAA Standard in VDM}}}.
\newblock In \bibinfo{editor}{S{\o}ren \surnamestart Prehn\surnameend} \&
  \bibinfo{editor}{W.~J. \surnamestart Toetenel\surnameend}, editors: {\sl
  \bibinfo{booktitle}{Formal Software Development -- Proceedings (Volume 1) of
  the 4th International Symposium of VDM Europe (VDM'91), Noordwijkerhout, The
  Netherlands}}, {\sl \bibinfo{series}{Lecture Notes in Computer Science}}
  \bibinfo{volume}{551}, \bibinfo{publisher}{Springer}, pp.
  \bibinfo{pages}{526--544}, \doi{10.1007/3-540-54834-3_31}.

\bibitemdeclare{incollection}{Preneel-11}
\bibitem{Preneel-11}
\bibinfo{author}{Bart \surnamestart Preneel\surnameend} (\bibinfo{year}{2011}):
  \emph{\bibinfo{title}{{MAA}}}.
\newblock In \bibinfo{editor}{Henk C.~A. \surnamestart van Tilborg\surnameend}
  \& \bibinfo{editor}{Sushil \surnamestart Jajodia\surnameend}, editors: {\sl
  \bibinfo{booktitle}{Encyclopedia of Cryptography and Security (2nd
  Edition)}}, \bibinfo{publisher}{Springer}, pp. \bibinfo{pages}{741--742},
  \doi{10.1007/978-1-4419-5906-5_591}.

\bibitemdeclare{inproceedings}{Preneel-vanOorschot-96}
\bibitem{Preneel-vanOorschot-96}
\bibinfo{author}{Bart \surnamestart Preneel\surnameend} \&
  \bibinfo{author}{Paul~C. \surnamestart van Oorschot\surnameend}
  (\bibinfo{year}{1996}): \emph{\bibinfo{title}{{On the Security of Two MAC
  Algorithms}}}.
\newblock In \bibinfo{editor}{Ueli~M. \surnamestart Maurer\surnameend}, editor:
  {\sl \bibinfo{booktitle}{Advances in Cryptology -- Proceedings of the
  International Conference on the Theory and Application of Cryptographic
  Techniques (EUROCRYPT'96), Saragossa, Spain}}, {\sl \bibinfo{series}{Lecture
  Notes in Computer Science}} \bibinfo{volume}{1070},
  \bibinfo{publisher}{Springer}, pp. \bibinfo{pages}{19--32},
  \doi{10.1007/3-540-68339-9_3}.

\bibitemdeclare{article}{Preneel-vanOorschot-99}
\bibitem{Preneel-vanOorschot-99}
\bibinfo{author}{Bart \surnamestart Preneel\surnameend} \&
  \bibinfo{author}{Paul~C. \surnamestart van Oorschot\surnameend}
  (\bibinfo{year}{1999}): \emph{\bibinfo{title}{{On the Security of Iterated
  Message Authentication Codes}}}.
\newblock {\sl \bibinfo{journal}{{IEEE} Transactions on Information Theory}}
  \bibinfo{volume}{45}(\bibinfo{number}{1}), pp. \bibinfo{pages}{188--199},
  \doi{10.1109/18.746787}.

\bibitemdeclare{article}{Preneel-Rumen-vanOorschot-97}
\bibitem{Preneel-Rumen-vanOorschot-97}
\bibinfo{author}{Bart \surnamestart Preneel\surnameend},
  \bibinfo{author}{Vincent \surnamestart Rumen\surnameend} \&
  \bibinfo{author}{Paul~C. \surnamestart van Oorschot\surnameend}
  (\bibinfo{year}{1997}): \emph{\bibinfo{title}{{Security Analysis of the
  Message Authenticator Algorithm (MAA)}}}.
\newblock {\sl \bibinfo{journal}{European Transactions on Telecommunications}}
  \bibinfo{volume}{8}(\bibinfo{number}{5}), pp. \bibinfo{pages}{455--470},
  \doi{10.1002/ett.4460080504}.

\bibitemdeclare{inproceedings}{Rijmen-Preneel-DeWin-96}
\bibitem{Rijmen-Preneel-DeWin-96}
\bibinfo{author}{Vincent \surnamestart Rijmen\surnameend},
  \bibinfo{author}{Bart \surnamestart Preneel\surnameend} \&
  \bibinfo{author}{Erik \surnamestart {De Win}\surnameend}
  (\bibinfo{year}{1996}): \emph{\bibinfo{title}{{Key Recovery and Collision
  Clusters for MAA}}}.
\newblock In: {\sl \bibinfo{booktitle}{Proceedings of the 1st International
  Conference on Security in Communication Networks (SCN'96)}}.
\newblock
  \urlprefix\url{https://www.cosic.esat.kuleuven.be/publications/article-437.p%
df}.

\end{thebibliography}
\end{document}